\documentclass{JHEP3} 

\usepackage{epsfig,multicol,bbm}
\usepackage{cite}
\usepackage{multirow}

\newcommand\fverb{\setbox\fverbbox=\hbox\bgroup\verb}
\newcommand\fverbdo{\egroup\medskip\noindent%
            \fbox{\unhbox\fverbbox}\ }
\newcommand\fverbit{\egroup\item[\fbox{\unhbox\fverbbox}]}
\newbox\fverbbox

\title{ NLO QCD + NLO EW corrections to $WZZ$ productions with leptonic decays at the LHC }

\author{Shen Yong-Bai,$^a$ Zhang Ren-You,$^a$ Ma Wen-Gan,$^a$ Li Xiao-Zhou,$^a$ Zhang Yu$^a$ and Guo Lei$^b$ \\
$^a$ Department of Modern Physics, University of Science and Technology of China, \\
~~\,96 Jinzhai Road, Hefei, Anhui 230026, P.R. China \\
$^b$ Department of Physics, Chongqing University, \\
~~\,55 Daxuecheng South Road, Chongqing 401331, P.R. China \\
E-mail: \email{ybshen@mail.ustc.edu.cn},
        \email{zhangry@ustc.edu.cn},
        \email{mawg@ustc.edu.cn},
        \email{lixz0818@mail.ustc.edu.cn},
        \email{dayu@mail.ustc.edu.cn},
        \email{guoleicqu@cqu.edu.cn} }

\abstract{
Precision tests of the Standard Model (SM) require not only accurate experiments, but also precise and reliable theoretical predictions. Triple vector boson production provides a unique opportunity to investigate the quartic gauge couplings and check the validity of the gauge principle in the SM. Since the tree-level predictions alone are inadequate to meet this demand, the next-to-leading order (NLO) calculation becomes compulsory. In this paper, we calculate the NLO QCD + NLO electroweak (EW) corrections to the $W^{\pm}ZZ$ productions with subsequent leptonic decays at the $14~{\rm TeV}$ LHC by adopting an improved narrow width approximation which takes into account the off-shell contributions and spin correlations from the $W^{\pm}$- and $Z$-boson leptonic decays. The NLO QCD+EW corrected integrated cross sections for the $W^{\pm}ZZ$ productions and some kinematic distributions of final products are provided. The results show that both the NLO QCD and NLO EW corrections are significant. In the jet-veto event selection scheme with $p_{T,jet}^{cut} = 50~ {\rm GeV}$, the NLO QCD+EW relative corrections to the integrated cross section are $20.5\%$ and $31.1\%$, while the genuine NLO EW relative corrections are $-5.42\%$ and $-4.58\%$, for the $W^+ZZ$ and $W^-ZZ$ productions, respectively. We also investigate the theoretical dependence of the integrated cross section on the factorization/renormalization scale, and find that the scale uncertainty is underestimated at the LO due to the fact that the strong coupling $\alpha_s$ is not involved in the LO matrix elements. }


\vfill \eject

\begin{document}

\vskip 5mm
\section{Introduction}
\par
After the discovery of the $126~ {\rm GeV}$ Higgs boson by the ATLAS and CMS collaborations \cite{126Higgs-1, 126Higgs-2}, the main tasks of further experiments at the Large Hadron Collider (LHC) are to determine the Higgs properties, test the predictions of the Standard Model (SM), and search for new physics beyond the SM. One of the important experiments for testing the SM is to measure the gauge couplings in the SM and check the validity of the gauge principle. The multiple gauge boson productions at the LHC can be used to determine the gauge boson self-couplings and help us for better understanding the electroweak (EW) symmetry breaking. The theoretical predictions for most multiple gauge boson productions at the LHC have been computed up to the QCD next-to-leading order (NLO) so far. The NLO EW corrections to most of these processes are not yet studied, although they are certainly significant in some cases. Therefore, precision theoretical predictions including both the NLO QCD and NLO EW corrections for the multiple gauge boson productions are necessary.

\par
The triple gauge boson productions are of particular interest because they are sensitive to both the triple and quartic gauge couplings (TGCs and QGCs) and thus related to the electroweak symmetry breaking mechanism \cite{EWSB-1, EWSB-2}. Therefore, the measurements of the triple gauge boson productions at hadron colliders can provide rich information on the gauge self-interactions and play an important role in searching for new physics beyond the SM. All the triple gauge boson productions at hadron colliders, $pp \rightarrow WWZ$, $ZZZ$, $WWW$, $WZZ$, $WW\gamma$, $ZZ\gamma$, $Z\gamma\gamma$, $\gamma\gamma\gamma$, $W\gamma\gamma$ and $WZ\gamma$, have been studied in the SM up to the QCD NLO \cite{WWZ, ZZZ, WWW-WZZ, WWZ+ZZZ+WWW+WZZ, WWr-ZZr, Zrr-rrr, Wrr, Wrr-decay, WZr}, while only the NLO EW correction to the $pp \to WWZ$ process has been calculated \cite{EWWWZ}. Therefore, the precision study on the $VV^{\prime}V^{\prime\prime}~(V,V^{\prime},V^{\prime\prime} = W~{\rm or}~Z)$ productions at hadron colliders with subsequent vector boson decays including the NLO QCD + NLO EW corrections is desired, and is listed in the Les Houches 2013 high precision wish list \cite{wishlist}.

\par
The $WZZ$ production at the LHC is sensitive to both the triple $WWZ$ coupling and quartic $WWZZ$ coupling and thus relevant for studying anomalous gauge couplings \cite{anomalous-coupling-1, anomalous-coupling-2}, and this production process with leptonic decays may serve as SM background in searching for new physics beyond the SM. The NLO QCD correction to the $WZZ$ production at the LHC was calculated in Refs.\cite{WWW-WZZ} and \cite{WWZ+ZZZ+WWW+WZZ}, while the NLO EW correction to this process with subsequent $W$- and $Z$-boson leptonic decays has not been investigated so far. In this paper we study the NLO QCD + NLO EW corrections to the $WZZ$ production with subsequent vector boson decays at the LHC. The rest of this paper is organized as follows: In section \ref{sec-calculations} we provide the details of our calculation strategy. The integrated cross sections and some kinematic distributions for the $pp \rightarrow WZZ + X$ process up to the QCD and EW NLO are presented and discussed in section \ref{sec-nresults}. Finally, a short summary is given in section \ref{sec-summary}.

\vskip 5mm
\section{Calculations}
\label{sec-calculations}
\par
We only take into account the Cabibbo-Kobayashi-Maskawa (CKM) mixing between the first two quark generations since the mixing to the third generation is negligible, i.e., the CKM matrix is $2 \oplus 1$ block-diagonal. The masses of the first two generations of quarks are set to zero. In this approximation, the CKM matrix factorizes from all the amplitudes, including the tree-level amplitudes for $WZZ$, $WZZ + g$, $WZZ + \gamma$, $WZZ + q$ productions and the QCD and EW one-loop amplitudes for $WZZ$ production. Therefore, only one generic amplitude for each  category mentioned above has to be evaluated when convoluting the squared matrix elements with the parton distribution functions (PDFs) \cite{Wr}. We adopt the 't Hooft-Feynman gauge and the four-flavor scheme in the calculations for the LO and NLO QCD + NLO EW corrections.

\par
\subsection{LO calculation }
\par
The LO contributions to the $W^+ZZ$ and $W^-ZZ$ productions at the LHC come from the following partonic processes:
\begin{eqnarray}
&& q_1(p_1) + \bar{q}_2(p_2) \to W^+(p_3) + Z (p_4) + Z (p_5) , \nonumber \\
&& \bar{q}_1(p_1) + q_2(p_2) \to W^-(p_3) + Z (p_4) + Z (p_5) , ~~~~~(q_1 = u, c,~ q_2 = d, s),
\end{eqnarray}
respectively. The parton-level cross section for the $q_2\bar{q}_1 \to W^-ZZ$ process in the SM should be the same as for the $q_1\bar{q}_2 \to W^+ZZ$ process due to the CP conservation. Therefore, we describe the LO and NLO calculations only for the $W^+ZZ$ production in this section.

\par
\begin{figure}[htbp]
  \begin{center}
     \includegraphics[scale=0.75]{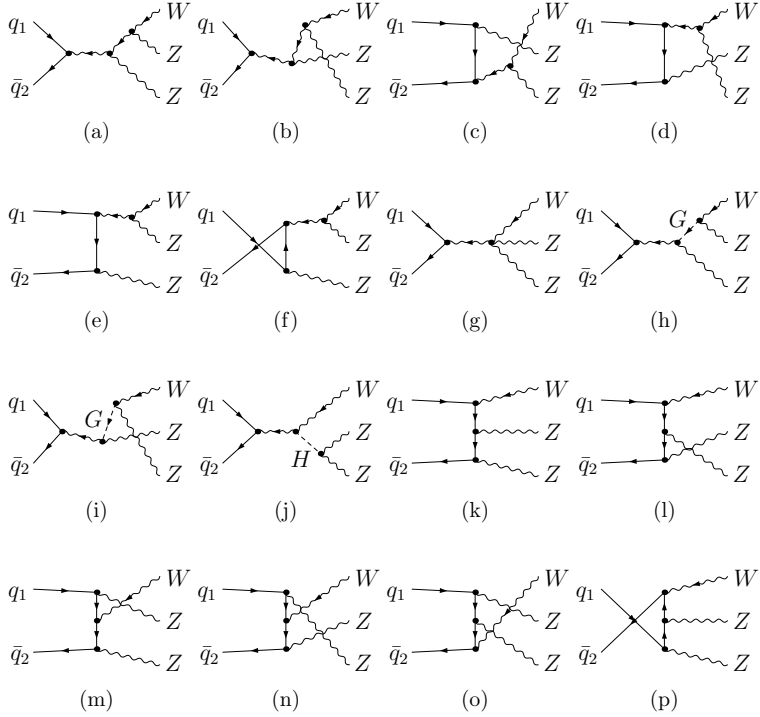}
    \caption{\small The LO Feynman diagrams for the $q_1\bar{q}_2 \rightarrow W^+ZZ$ partonic process, where $H$ and $G$ represent the Higgs and charged Goldstone bosons, respectively.}
   \label{fig:born}
  \end{center}
\end{figure}

The LO Feynman diagrams for the $q_1\bar{q}_2 \to W^+ZZ$ partonic process are shown in Fig.\ref{fig:born}. The $WWZ$ TGC is involved in Figs.\ref{fig:born}(a,b,c,d,e,f) and only Fig.\ref{fig:born}(g) contains the $WWZZ$ QGC. The LO parton-level cross section for $q_1\bar{q}_2 \to W^+ZZ$ is expressed as
\begin{eqnarray}
\label{losigma}
\hat{\sigma}_{LO}^{q_1\bar{q}_2}(\hat{s})
=
\frac{1}{2} \frac{1}{2\hat{s}} \int \overline{\sum}
\left|{\cal M}^{(0)}_{q_1\bar{q}_2}\right|^2
d\Omega_3,
\end{eqnarray}
where the factor $\frac{1}{2}$ arises from the two identical $Z$-bosons in the final state. The summation is taken over the spins of the final state, and the bar over the summation represents averaging over the spins and colors of the initial state. ${\cal M}^{(0)}_{q_1\bar{q}_2}$ is the LO Feynman amplitude for the $q_1\bar{q}_2 \to W^+ZZ$ partonic process, and $d\Omega_3$ is the three-body final state phase space element defined as
\begin{eqnarray}
d\Omega_3
=
(2 \pi)^4 \delta^{(4)}(p_1+p_2-p_3-p_4-p_5)
\frac{d^3 \vec{p}_3}{(2\pi)^3 2E_3}
\frac{d^3 \vec{p}_4}{(2\pi)^3 2E_4}
\frac{d^3 \vec{p}_5}{(2\pi)^3 2E_5}.
\end{eqnarray}

\par
The final produced $W^+$- and $Z$-bosons are unstable particles, and we only consider their leptonic decay modes in investigating the $W^+ZZ$ production. Then the LO cross section for the $pp \rightarrow W^+ZZ \rightarrow \ell^+_1 \nu_{\ell_1} \ell^+_2 \ell^-_2 \ell^+_3 \ell^-_3 + X$ process can be obtained by performing the PDF convolution and applying the narrow width approximation as
\begin{eqnarray}
&&\sigma_{LO}(pp \rightarrow W^+ZZ \rightarrow \ell^+_1 \nu_{\ell_1} \ell^+_2 \ell^-_2 \ell^+_3 \ell^-_3 + X, S) \nonumber \\
&=&
\sum_{q_1 = u,c}^{q_2 = d,s}
\int_0^1 dx_1 dx_2
\Big[
\Phi_{q_1|P_1}(x_1, \mu_F) \Phi_{\bar{q}_2|P_2}(x_2, \mu_F) + \left( P_1 \leftrightarrow P_2 \right)
\Big]
\hat{\sigma}_{LO}^{q_1\bar{q}_2}(\hat{s} = x_1x_2S) \nonumber \\
&& \times
\Big[ \sum_{\ell=e, \mu, \tau}Br(W^+ \rightarrow \ell^+\nu_{\ell}) \Big]
\Big[ \sum_{\ell=e, \mu, \tau}Br(Z \rightarrow \ell^+\ell^-) \Big]^2,
\end{eqnarray}
where $\Phi_{q|P}$ is the quark density in proton, $\mu_F$ is the factorization scale, and $\sqrt{\hat{s}}$ and $\sqrt{S}$ are the center-of-mass system energies of the initial $q_1\bar{q}_2$ quark pair and colliding protons, respectively.

\par
\subsection{NLO QCD corrections}
\label{sec-NLOQCD}
\par
The NLO QCD correction to the parent process $pp \rightarrow W^+ZZ + X$ includes: (1) The QCD one-loop virtual corrections to $q_1\bar{q}_2 \rightarrow W^+ZZ$; (2) The real gluon emission corrections from $q_1\bar{q}_2 \rightarrow W^+ZZ + g$; (3) The real light-quark emission corrections from $q_1 g \rightarrow W^+ZZ + q_2$ and $\bar{q}_2 g \rightarrow W^+ZZ + \bar{q}_1$, (where $q_1 = u, c,~ q_2 = d, s$); (4) The corresponding contributions of the PDF counterterms.

\par
An ultraviolet (UV) and infrared (IR) safe observable requires the exact cancelation of the UV and IR singularities to all orders. In this paper, we adopt the dimensional regularization scheme in $D=4-2\epsilon$ dimensions to isolate both the UV and IR singularities. After renormalizing the related quark wave functions $\psi_{q_1}$ and $\psi_{q_2}$, the QCD one-loop virtual correction to the $q_1\bar{q}_2 \rightarrow W^+ZZ$ partonic process is UV-finite but still contains soft and collinear IR singularities. The soft IR singularity is canceled exactly by that in the real gluon emission correction from the $q_1\bar{q}_2 \rightarrow W^+ZZ + g$ partonic process, while the collinear IR singularity is only partially canceled by that in the real gluon emission correction from $q_1\bar{q}_2 \rightarrow W^+ZZ + g$ and the remaining collinear IR singularity is absorbed by the collinear gluon emission parts of the related quark PDF QCD counterterms. The real light-quark emission corrections from $q_1 g \rightarrow W^+ZZ + q_2$ and $\bar{q}_2 g \rightarrow W^+ZZ + \bar{q}_1$ only contain collinear IR singularities, and can be canceled exactly by the collinear quark emission parts of the PDF QCD counterterms for $q_1$ and $\bar{q}_2$, respectively. Therefore, $\Delta \sigma^{q_1\bar{q}_2}_{QCD}$, $\Delta \sigma^{q_1g}_{QCD}$ and $\Delta \sigma^{\bar{q}_2g}_{QCD}$, defined as
\begin{eqnarray}
\label{QCD correction parts}
\Delta \sigma^{q_1\bar{q}_2}_{QCD}
&=&
\int dx_1 dx_2
\Big[
\Phi_{q_1|P_1}(x_1, \mu_F) \Phi_{\bar{q}_2|P_2}(x_2, \mu_F) + \left( P_1 \leftrightarrow P_2, x_1 \leftrightarrow x_2 \right)
\Big]
\left(
\hat{\sigma}_{QCD}^{q_1\bar{q}_2,V} + \hat{\sigma}_{QCD}^{q_1\bar{q}_2,R}
\right) \nonumber \\
&+&
\int dx_1 dx_2
\Big[
\delta \Phi^{QCD,(g)}_{q_1|P_1}(x_1, \mu_F) \Phi_{\bar{q}_2|P_2}(x_2, \mu_F)
+
\Phi_{q_1|P_1}(x_1, \mu_F) \delta \Phi^{QCD,(g)}_{\bar{q}_2|P_2}(x_2, \mu_F) \nonumber \\
&&
~~~~~~~~~~~~~ + \left( P_1 \leftrightarrow P_2, x_1 \leftrightarrow x_2 \right)
\Big]
\hat{\sigma}_{LO}^{q_1\bar{q}_2}, \nonumber \\
\Delta \sigma^{q_1g}_{QCD}
&=&
\int dx_1 dx_2
\Big[
\Phi_{q_1|P_1}(x_1, \mu_F) \Phi_{g|P_2}(x_2, \mu_F) + \left( P_1 \leftrightarrow P_2, x_1 \leftrightarrow x_2 \right)
\Big]
\hat{\sigma}_{QCD}^{q_1g,R} \nonumber \\
&+&
\int dx_1 dx_2
\Big[
\Phi_{q_1|P_1}(x_1, \mu_F) \delta \Phi^{QCD,(q)}_{\bar{q}_2|P_2}(x_2, \mu_F) + \left( P_1 \leftrightarrow P_2, x_1 \leftrightarrow x_2 \right)
\Big]
\hat{\sigma}_{LO}^{q_1\bar{q}_2}, \nonumber \\
\Delta \sigma^{\bar{q}_2g}_{QCD}
&=&
\int dx_1 dx_2
\Big[
\Phi_{g|P_1}(x_1, \mu_F) \Phi_{\bar{q}_2|P_2}(x_2, \mu_F) + \left( P_1 \leftrightarrow P_2, x_1 \leftrightarrow x_2 \right)
\Big]
\hat{\sigma}_{QCD}^{\bar{q}_2g,R} \nonumber \\
&+&
\int dx_1 dx_2
\Big[
\delta \Phi^{QCD,(q)}_{q_1|P_1}(x_1, \mu_F) \Phi_{\bar{q}_2|P_2}(x_2, \mu_F) + \left( P_1 \leftrightarrow P_2, x_1 \leftrightarrow x_2 \right)
\Big]
\hat{\sigma}_{LO}^{q_1\bar{q}_2},
\end{eqnarray}
are both UV- and IR-finite separately, where $\hat{\sigma}_{QCD}^{q_1\bar{q}_2,V}$, $\hat{\sigma}_{QCD}^{q_1\bar{q}_2,R}$, $\hat{\sigma}_{QCD}^{q_1g,R}$ and $\hat{\sigma}_{QCD}^{\bar{q}_2g,R}$ are the NLO QCD virtual and real emission corrections from $q_1\bar{q}_2 \rightarrow W^+ZZ$, $q_1\bar{q}_2 \rightarrow W^+ZZ + g$, $q_1 g \rightarrow W^+ZZ + q_2$ and $\bar{q}_2 g \rightarrow W^+ZZ + \bar{q}_1$, respectively. The quark PDF QCD counterterm $\delta \Phi^{QCD}_{q|P}$ can be split into two parts: the collinear gluon emission part and collinear light-quark emission part, i.e., $\delta \Phi^{QCD}_{q|P} = \delta \Phi_{q|P}^{QCD,(g)} + \delta \Phi^{QCD,(q)}_{q|P}$, which can be expressed in the $\overline{MS}$ factorization scheme as
\begin{eqnarray}
&& \delta \Phi^{QCD,(g)}_{q|P}(x,\mu_F,\mu_R) =
   \frac{1}{\epsilon} \left[
                      \frac{C_F\alpha_s}{2 \pi}
                      \frac{\Gamma(1 - \epsilon)}{\Gamma(1 - 2 \epsilon)}
                      \left( \frac{4 \pi \mu_R^2}{\mu_F^2} \right)^{\epsilon}
                      \right]
   \int_x^1 \frac{dz}{z} \left[P_{qq}(z)\right]_+ \Phi_{q|P}(x/z,\mu_F), \nonumber \\
&& \delta \Phi^{QCD,(q)}_{q|P}(x,\mu_F,\mu_R) =
   \frac{1}{\epsilon} \left[
                      \frac{T_F\alpha_s}{2 \pi}
                      \frac{\Gamma(1 - \epsilon)}{\Gamma(1 - 2 \epsilon)}
                      \left( \frac{4 \pi \mu_R^2}{\mu_F^2} \right)^{\epsilon}
                      \right]
   \int_x^1 \frac{dz}{z} P_{qg}(z) \Phi_{g|P}(x/z,\mu_F),~~~~~~~
\end{eqnarray}
where $C_F = 4/3$, $T_F=1/2$ and $\mu_R$ is the renormalization scale. The splitting functions are given by
\begin{eqnarray}
\label{eq:spliting}
P_{qq}(z) = \frac{1 + z^2}{1 - z}, ~~~~~ P_{qg}(z) = z^2 + (1 - z)^2,
\end{eqnarray}
and the $\left[ \ldots \right]_+$ prescription is understood as
\begin{eqnarray}
\int_0^1 dz \left[ g(z) \right]_+ f(z) = \int_0^1 dz \, g(z) \left[ f(z) - f(1) \right].
\end{eqnarray}

\par
In our NLO calculation we adopt the two cutoff phase space slicing (TCPSS) technique to isolate the IR singularities for the real emission partonic processes \cite{tcpss}. Two cutoffs $\delta_{s}$ and $\delta_{c}$ are introduced to separate the phase space of the real gluon emission process $q_1(p_1) + \bar{q}_2(p_2) \rightarrow W^+(p_3) + Z(p_4) + Z(p_5) + g(p_6)$ into soft gluon region ($E_{6} \leq \delta_{s} \sqrt{\hat{s}}/2$), hard collinear region ($E_{6} > \delta_{s} \sqrt{\hat{s}}/2$ and $\min\{ \hat{s}_{16}, \hat{s}_{26}\} \leq \delta_{c}\hat{s}$) and hard noncollinear region ($E_{6} > \delta_{s} \sqrt{\hat{s}}/2$ and $\min\{ \hat{s}_{16}, \hat{s}_{26}\} > \delta_{c}\hat{s}$). The real light-quark emission processes, $q_1(p_1) + g(p_2) \rightarrow W^+(p_3) + Z(p_4) + Z(p_5) + q_2(p_6)$ and $\bar{q}_2(p_1) + g(p_2) \rightarrow W^+(p_3) + Z(p_4) + Z(p_5) + \bar{q}_1(p_6)$, contain only collinear IR singularities, therefore we only separate their phase space into collinear ($\hat{s}_{26} \leq \delta_{c}\hat{s}$) and noncollinear ($\hat{s}_{26} > \delta_{c}\hat{s}$) regions. Then we can express the cross sections for these real emission partonic processes as
\begin{eqnarray}
\label{QCD real corrections}
&& \hat{\sigma}_{QCD}^{q_1\bar{q}_2,R} = \hat{\sigma}_{QCD}^{q_1\bar{q}_2,S} + \hat{\sigma}_{QCD}^{q_1\bar{q}_2,HC} + \hat{\sigma}_{QCD}^{q_1\bar{q}_2,\overline{HC}}, \nonumber \\
&& \hat{\sigma}_{QCD}^{q_1g,R} = \hat{\sigma}_{QCD}^{q_1g,C} + \hat{\sigma}_{QCD}^{q_1g,\overline{C}},~~~~~~~~~
   \hat{\sigma}_{QCD}^{\bar{q}_2g,R} = \hat{\sigma}_{QCD}^{\bar{q}_2g,C} + \hat{\sigma}_{QCD}^{\bar{q}_2g,\overline{C}},
\end{eqnarray}
where the superscripts $S$, $HC$, $\overline{HC}$, $C$ and $\overline{C}$ stand for soft, hard collinear, hard noncollinear, collinear and noncollinear, respectively. The cross sections over the soft and (hard) collinear regions contain only the soft and collinear IR singularities separately, while the cross sections over the (hard) noncollinear regions are IR-finite. We have checked numerically the cutoff independence of the total cross sections for these subprocesses by setting $\delta_c = \delta_s/50$ and varying $\delta_s$ from $10^{-5}$ to $10^{-3}$.

\par
Finally, we get the full NLO QCD correction to the parent process $pp \rightarrow W^+ZZ + X$ as
\begin{eqnarray}
\label{pure QCD correction}
\Delta \sigma_{QCD}
=
\sum_{q_1=u,c}^{q_2=d,s}
\Big[
\Delta \sigma^{q_1\bar{q}_2}_{QCD}
+ \Delta \sigma^{q_1g}_{QCD}
+ \Delta \sigma^{\bar{q}_2g}_{QCD}
\Big].
\end{eqnarray}
We employ the modified FeynArts 3.7 package \cite{feynarts} to generate Feynman diagrams and their corresponding amplitudes. The reduction of output amplitudes are implemented by the FormCalc 7.3 package \cite{formcalc}. In our numerical calculation, the tensor 4-point integrals with rank $n>3$ may induce a serious unstable problem. We adopt our developed codes based on the LoopTools 2.8 package \cite{ff} to calculate the scaler and tensor integrals, which can switch to the quadruple precision arithmetic automatically if necessary.

\par
In Ref.\cite{WWW-WZZ} the authors took the same input parameters as in Ref.\cite{WWZ+ZZZ+WWW+WZZ} and made a numerical comparison for the $pp \to W^+ZZ+X$ process at the $\sqrt{S}=14~ {\rm TeV}$ LHC. We follow their input parameters and settings, and present in Table \ref{tab1} a comparison between our numerical results for the LO and NLO QCD corrected integrated cross sections and the corresponding ones provided in Ref.\cite{WWW-WZZ} and Ref.\cite{WWZ+ZZZ+WWW+WZZ}. It shows that all these numerical results are in good agreement with each other within the Monte Carlo errors.

\begin{table}[htbp]
\begin{center}
\small
\begin{tabular}{c|c|c}
\hline
$pp \rightarrow W^+ZZ+X$ & ~$\sigma_{LO}$ $[fb]$~ & ~$\sigma_{NLO}$ $[fb]$~  \\
\hline\hline
ours & 20.3(1) & 39.7(2)  \\
\hline  Ref.\cite{WWW-WZZ}         &20.24(3)                  &39.86(7)         \\
\hline  Ref.\cite{WWZ+ZZZ+WWW+WZZ}             &20.0(1)                   &39.7(2)          \\
\hline
\end{tabular}
\caption{\small  \label{tab1}
Comparison of our numerical results for the LO and NLO QCD corrected integrated cross sections with the corresponding ones in previous works \cite{WWW-WZZ,WWZ+ZZZ+WWW+WZZ}. All input parameters and settings are taken from Ref.\cite{WWW-WZZ} with $\mu_F = \mu_R = 3M_Z$.   }
\end{center}
\end{table}

\par
\subsection{NLO EW corrections}
\par
The NLO EW correction to the parent process $pp \rightarrow W^+ZZ + X$ includes: (1) The EW one-loop virtual corrections to $q_1\bar{q}_2 \rightarrow W^+ZZ$; (2) The real photon emission corrections from $q_1\bar{q}_2 \rightarrow W^+ZZ + \gamma$; (3) The contributions of the photon-induced partonic processes $q_1 \gamma \rightarrow W^+ZZ + q_2$ and $\bar{q}_2 \gamma \rightarrow W^+ZZ + \bar{q}_1$, (where $q_1 = u, c,~ q_2 = d, s$); (4) The corresponding contributions of the PDF counterterms.

\par
The UV divergences in the EW one-loop virtual corrections can be removed by the renormalization procedure. We take the definitions for the relevant EW renormalization constants same as in Ref.\cite{W-physics}. We adopt the on-mass-shell scheme to renormalize the masses and wave functions of related particles as used in the QCD correction. The expressions for the relevant renormalization constants and the unrenormalized EW self-energies can be found in Ref.\cite{W-physics}. In our calculation, we adopt a mixed scheme to deal with the EW couplings. All the EW couplings in the tree-level diagrams for $q_1\bar{q}_2 \rightarrow W^+ZZ$ are fixed in the $G_{\mu}$-scheme, i.e., $\alpha = \alpha_{G_{\mu}}$, while the extra EW couplings appeared in the EW one-loop diagrams for $q_1\bar{q}_2 \rightarrow W^+ZZ$ and in the real photon emission and photon-induced subprocesses are fixed in the $\alpha(0)$-scheme, i.e., $\alpha = \alpha(0)$ \footnote{The fine structure constant in the PDF EW counterterms (see Eq.(\ref{PDF-EWCT})) is also fixed as $\alpha = \alpha(0)$.}. $\alpha(0)$ is the fine structure constant in the Thomson limit and $\alpha_{G_{\mu}}$ is given by
\begin{eqnarray}
\label{alpha-G}
\alpha_{G_\mu} = \frac{\sqrt{2}}{\pi}G_{\mu}M_W^2\sin^2\theta_W
\end{eqnarray}
where $\sin^2 \theta_W = 1 - M_W^2/M_Z^2$. Then the LO cross section and NLO EW correction are of ${\cal O}(\alpha_{G_{\mu}}^3)$ and ${\cal O}(\alpha_{G_{\mu}}^3 \alpha(0))$, respectively, and correspondingly the electric charge renormalization constant should be given in the $G_{\mu}$-scheme as
\begin{eqnarray}
\delta Z_e^{G_\mu} = \delta Z_e^{\alpha(0)} -\frac{1}{2}\Delta r
= -\frac{1}{2}\delta Z_{AA} - \frac{1}{2} \tan\theta_W \delta Z_{ZA} - \frac{1}{2}\Delta r,
\end{eqnarray}
where $\delta Z_e^{\alpha(0)}$ is the electric charge renormalization constant in the $\alpha(0)$-scheme, and $\Delta r$ \cite{W-physics,W-production} corresponds to the subtraction of the logarithmic divergence contributed by the light quarks to $\delta Z_{AA}$ which was absorbed by $\alpha_{G_{\mu}}$.

\par
The quark PDF EW counterterm $\delta \Phi_{q|P}^{EW}$ also contains two parts: the collinear photon emission part and collinear light-quark emission part, i.e., $\delta \Phi^{EW}_{q|P} = \delta \Phi_{q|P}^{EW,(\gamma)} + \delta \Phi_{q|P}^{EW,(q)}$. In the DIS factorization scheme used for NLO EW corrections, these two collinear parts are expressed as \cite{EWWWZ,DIS-facscheme}
\begin{eqnarray}
\label{PDF-EWCT}
&& \delta \Phi_{q|P}^{EW,(\gamma)}(x,\mu_F,\mu_R) =
   \frac{Q_q^2\alpha}{2 \pi}
   \int_x^1 \frac{dz}{z}
   \Phi_{q|P}(x/z,\mu_F)
   \left\{
   \frac{1}{\epsilon}
                      \frac{\Gamma(1 - \epsilon)}{\Gamma(1 - 2 \epsilon)}
                      \left( \frac{4 \pi \mu_R^2}{\mu_F^2} \right)^{\epsilon}
   \left[ P_{qq}(z) \right]_+
 - C_{qq}^{DIS}(z)
   \right\}, \nonumber \\
&& \delta \Phi_{q|P}^{EW,(q)}(x,\mu_F,\mu_R) =
   \frac{3Q_q^2\alpha}{2 \pi}
   \int_x^1 \frac{dz}{z}
   \Phi_{\gamma|P}(x/z,\mu_F)
   \left\{
   \frac{1}{\epsilon}
                      \frac{\Gamma(1 - \epsilon)}{\Gamma(1 - 2 \epsilon)}
                      \left( \frac{4 \pi \mu_R^2}{\mu_F^2} \right)^{\epsilon}
   P_{q\gamma}(z)
 - C_{q\gamma}^{DIS}(z)
   \right\}, \nonumber \\
\end{eqnarray}
where $Q_q$ is the electric charge carried by the initial quark $q$, and $P_{q\gamma}(z) = P_{qg}(z)$. The DIS factorization scheme is specified by \cite{EWWWZ,DIS-facscheme}
\begin{eqnarray}
&& C_{qq}^{DIS}(z)
   =
   \left[
   P_{qq}(z) \left( \ln \frac{1-z}{z} - \frac{3}{4} \right) + \frac{9 + 5 z}{4} \right]_+ ,\nonumber \\
&& C_{q\gamma}^{DIS}(z)
   =
   P_{q\gamma}(z)
   \ln \frac{1-z}{z} - 8z^2 + 8z - 1.
\end{eqnarray}
Analogous to the QCD correction described in Sec.\ref{sec-NLOQCD}, we get the full NLO EW correction to the parent process $pp \rightarrow W^+ZZ + X$ as
\begin{eqnarray}
\label{full EW correction}
\Delta \sigma_{EW}
=
\sum_{q_1=u,c}^{q_2=d,s}
\Big[
\Delta \sigma^{q_1\bar{q}_2}_{EW}
+ \Delta \sigma^{q_1\gamma}_{EW}
+ \Delta \sigma^{\bar{q}_2\gamma}_{EW}
\Big],
\end{eqnarray}
where the analytic expressions for $\Delta \sigma^{q_1\bar{q}_2}_{EW}$, $\Delta \sigma^{q_1\gamma}_{EW}$ and $\Delta \sigma^{\bar{q}_2\gamma}_{EW}$ can be obtained from Eqs.(\ref{QCD correction parts}) and (\ref{QCD real corrections}) by doing the replacements of $QCD \rightarrow EW$ and $g \rightarrow \gamma$. Each of $\Delta \sigma^{q_1\bar{q}_2}_{EW}$, $\Delta \sigma^{q_1\gamma}_{EW}$ and $\Delta \sigma^{\bar{q}_2\gamma}_{EW}$ ($q_1 = u, c,~ q_2 = d, s$) is UV- and IR-finite, therefore the full NLO EW correction $\Delta \sigma_{EW}$ is an UV- and IR-safe variable.

\subsection{CKM matrix dependence}
\par
The $W^+ZZ$ production at the LHC up to the QCD and EW NLO involves the following topologies:
\begin{eqnarray}
\label{CKMfac-M}
& 0 \rightarrow W^+ ZZ + \bar{q}_1 q_2\,\,\,\,\,\,\,\,\,\, & ~~~~~({\rm tree,~ QCD~ and~ EW~ loop}), \nonumber \\
& 0 \rightarrow W^+ ZZ + \bar{q}_1 q_2 + g &  ~~~~~({\rm tree}),  \nonumber \\
& 0 \rightarrow W^+ ZZ + \bar{q}_1 q_2 + \gamma & ~~~~~({\rm tree}),
\end{eqnarray}
where $q_1 = u, c$ and $q_2 = d, s$. Each of these topologies contains only one charged current quark chain,\footnote{We do not consider the closed quark loop, because the CKM matrix in it drops out after the summation over quark flavors.} therefore the CKM matrix element can factorize from the amplitudes as
\begin{eqnarray}
{\cal M}_{{\rm tree,loop}}(0 \rightarrow W^+ ZZ + \bar{q}_1 q_2) &=& V_{q_1 q_2}^{\ast} \times \widetilde{{\cal M}}_{{\rm tree,loop}}(0 \rightarrow W^+ ZZ + \bar{u} d), \nonumber \\
{\cal M}_{{\rm tree}}(0 \rightarrow W^+ ZZ + \bar{q}_1 q_2 + g) &=& V_{q_1 q_2}^{\ast} \times \widetilde{{\cal M}}_{{\rm tree}}(0 \rightarrow W^+ ZZ + \bar{u} d + g), \nonumber \\
{\cal M}_{{\rm tree}}(0 \rightarrow W^+ ZZ + \bar{q}_1 q_2 + \gamma) &=& V_{q_1 q_2}^{\ast} \times \widetilde{{\cal M}}_{{\rm tree}}(0 \rightarrow W^+ ZZ + \bar{u} d + \gamma),
\end{eqnarray}
since $m_u=m_c=m_d = m_s=0$, where ${\cal M}$ and $\widetilde{{\cal M}}$ are the amplitudes obtained with and without CKM matrix.

\par
For convenience we introduce the following CKM matrix dependent structure functions:
\begin{eqnarray}
{\cal F}_{q\bar{q}}(x_1, x_2, \mu_F)
&=&
\sum_{q_1 = u,c}^{q_2 = d,s}
\left|V_{q_1 q_2}\right|^2 \Big[ \Phi_{q_1|P_1}(x_1, \mu_F) \Phi_{\bar{q}_2|P_2}(x_2, \mu_F)
+ \left( P_1 \leftrightarrow P_2, x_1 \leftrightarrow x_2 \right) \Big], \nonumber \\
{\cal F}_{qg}(x_1, x_2, \mu_F)
&=&
\sum_{q_1 = u,c}^{q_2 = d,s}
\left|V_{q_1 q_2}\right|^2 \Big[ \Phi_{q_1|P_1}(x_1, \mu_F) \Phi_{g|P_2}(x_2, \mu_F)
+ \left( P_1 \leftrightarrow P_2, x_1 \leftrightarrow x_2 \right) \Big], \nonumber \\
{\cal F}_{\bar{q}g}(x_1, x_2, \mu_F)
&=&
\sum_{q_1 = u,c}^{q_2 = d,s}
\left|V_{q_1 q_2}\right|^2 \Big[ \Phi_{g|P_1}(x_1, \mu_F) \Phi_{\bar{q}_2|P_2}(x_2, \mu_F)
+ \left( P_1 \leftrightarrow P_2, x_1 \leftrightarrow x_2 \right) \Big], \nonumber \\
{\cal F}_{q\gamma}(x_1, x_2, \mu_F)
&=&
{\cal F}_{qg}(x_1, x_2, \mu_F)\Big|_{g \rightarrow \gamma}, \nonumber \\
{\cal F}_{\bar{q}\gamma}(x_1, x_2, \mu_F)
&=&
{\cal F}_{\bar{q}g}(x_1, x_2, \mu_F)\Big|_{g \rightarrow \gamma}.
\end{eqnarray}
Then the full NLO QCD and EW corrections to the parent process $pp \rightarrow W^+ZZ + X$ can be simply rewritten as
\begin{eqnarray}
\label{NLO corrections-F}
\Delta \sigma_{QCD}
&=&
\int dx_1 dx_2
\left[
{\cal F}_{q\bar{q}} \left( \hat{\tilde{\sigma}}_{QCD}^{u\bar{d}, V} + \hat{\tilde{\sigma}}_{QCD}^{u\bar{d}, R} \right)
+ {\cal F}_{qg} \hat{\tilde{\sigma}}_{QCD}^{ug, R}
+ {\cal F}_{\bar{q}g} \hat{\tilde{\sigma}}_{QCD}^{\bar{d}g, R}
+ \delta {\cal F}_{q\bar{q}}^{QCD} \hat{\tilde{\sigma}}_{LO}^{u\bar{d}}
\right], ~~~ \nonumber \\
\Delta \sigma_{EW}
&=&
\int dx_1 dx_2
\left[
{\cal F}_{q\bar{q}} \left( \hat{\tilde{\sigma}}_{EW}^{u\bar{d}, V} + \hat{\tilde{\sigma}}_{EW}^{u\bar{d}, R} \right)
+ {\cal F}_{q\gamma} \hat{\tilde{\sigma}}_{EW}^{u\gamma, R}
+ {\cal F}_{\bar{q}\gamma} \hat{\tilde{\sigma}}_{EW}^{\bar{d}\gamma, R}
+ \delta {\cal F}_{q\bar{q}}^{EW} \hat{\tilde{\sigma}}_{LO}^{u\bar{d}}
\right],
\end{eqnarray}
where $\delta {\cal F}_{q\bar{q}}^{QCD}$ and $\delta {\cal F}_{q\bar{q}}^{EW}$ are the QCD and EW counterterms of ${\cal F}_{q\bar{q}}$, respectively, expressed as
\begin{eqnarray}
\delta {\cal F}_{q\bar{q}}^{QCD}
&=&
\sum_{q_1 = u,c}^{q_2 = d,s}
\left|V_{q_1 q_2}\right|^2
\Big[
\delta \Phi_{q_1|P_1}^{QCD}(x_1, \mu_F) \Phi_{\bar{q}_2|P_2}(x_2, \mu_F) + \Phi_{q_1|P_1}(x_1, \mu_F) \delta \Phi_{\bar{q}_2|P_2}^{QCD}(x_2, \mu_F) \nonumber \\
&& ~~~~~~~~~~~~~~~~~~
+ \left( P_1 \leftrightarrow P_2, x_1 \leftrightarrow x_2 \right)
\Big], \nonumber \\
\delta {\cal F}_{q\bar{q}}^{EW}
&=&
\delta {\cal F}_{q\bar{q}}^{QCD}\Big|_{QCD \rightarrow EW},
\end{eqnarray}
and $\hat{\tilde{\sigma}}_{LO}^{u\bar{d}}$, $\hat{\tilde{\sigma}}_{QCD}^{u\bar{d}, V}$, $\hat{\tilde{\sigma}}_{QCD}^{u\bar{d}, R}$, $\hat{\tilde{\sigma}}_{QCD}^{ug, R}$, $\hat{\tilde{\sigma}}_{QCD}^{\bar{d}g, R}$, $\hat{\tilde{\sigma}}_{EW}^{u\bar{d}, V}$, $\hat{\tilde{\sigma}}_{EW}^{u\bar{d}, R}$, $\hat{\tilde{\sigma}}_{EW}^{u\gamma, R}$, $\hat{\tilde{\sigma}}_{EW}^{\bar{d}\gamma, R}$ are the corresponding partonic cross sections by setting $V_{CKM} = I$. From Eq.(\ref{NLO corrections-F}) we see clearly that the CKM matrix factorizes from the amplitudes and is absorbed by the structure functions. Therefore, only the amplitudes for one generation of quarks have to be evaluated in PDF convolution \footnote{In the case of $V_{CKM} = I$ and $m_u=m_c=m_d = m_s=0$, the amplitudes for the first quark generation are the same as the corresponding ones for the second quark generation. In this paper we compute the related amplitudes and partonic cross sections only for the first generation of quarks (see Eqs.(\ref{CKMfac-M}) and (\ref{NLO corrections-F})).}.

\par
\section{Numerical results and discussion}
\label{sec-nresults}
\par
\subsection{Input parameters}
\par
The SM input parameters are taken as \cite{PDG}:
\begin{eqnarray}
&& M_W = 80.385~{\rm GeV},~~ M_Z = 91.1876~{\rm GeV},~~ M_H = 126~{\rm GeV},~~ m_t=173.5~{\rm GeV}, \nonumber \\
&& G_{\mu} = 1.16638\times10^{-5}~{\rm GeV}^{-2},~~ \alpha(0) = 1/137.036,~~ \alpha_s(m_Z) = 0.119.~~~~
\end{eqnarray}
The masses of all leptons and quarks but the top quark are set to zero.
The CKM matrix elements are taken as
\begin{eqnarray}\label{CKM}
 V_{CKM} &=& \left(
\begin{array}{ccc}
    V_{ud} \ &  V_{us} \ &  V_{ub} \\
    V_{cd} \ &  V_{cs} \ &  V_{cb} \\
    V_{td} \ &  V_{ts} \ &  V_{tb} \\
\end{array}
    \right)=\left(
\begin{array}{ccc}
     0.97425 \ &  0.22547 \ &  0 \\
    -0.22547 \ &  0.97425 \ &  0 \\
       0 \ &  0 \ &  1
\end{array}  \right).
\end{eqnarray}
We set the factorization and renormalization scales being equal and choose the central scale as $\mu_0 = \frac{1}{2} M_W + M_Z$. The two cutoffs are set as $\delta_s = 10^{-3}$, $\delta_c = 2 \times 10^{-5}$ in NLO QCD calculation, and $\delta_s = 10^{-4}$, $\delta_c = 2 \times 10^{-6}$ in NLO EW calculation, respectively.

\par
In the calculation of the NLO QCD corrections, we adopt the NLO NNPDF2.3QED PDFs \cite{NNPDF} with the $\overline{MS}$ factorization scheme. The strong coupling constant $\alpha_s$ is renormalized in the $\overline{MS}$ scheme with five active flavors, and its running value $\alpha_s(\mu)$ is taken from the PDF set. While for the NLO EW corrections we use the  NLO NNPDF2.3QED PDFs with the DIS factorization scheme \cite{DIS-facscheme}.

\par
\subsection{Integrated cross sections}
\par
The NLO QCD and EW relative corrections are defined as
\begin{eqnarray}
\label{relative correction}
\delta_{QCD}
=
\frac{\Delta \sigma_{QCD} + \left( \sigma_{0} - \sigma_{LO} \right)}{\sigma_{LO}},~~~~~~~~
\delta_{EW}
=
\frac{\Delta \sigma_{EW}}{\sigma_{0}},
\end{eqnarray}
where $\Delta \sigma_{QCD}$ and $\Delta \sigma_{EW}$ (see Eqs.(\ref{pure QCD correction}) and (\ref{full EW correction})) are evaluated with NLO PDFs, and $\sigma_{LO}$ and $\sigma_{0}$ are LO cross sections calculated with LO and NLO PDFs, respectively. The numerator $\Delta \sigma_{QCD} + \left( \sigma_{0} - \sigma_{LO} \right)$ represents the full NLO QCD correction that includes all the NLO QCD contributions from both the dynamic matrix element and PDFs. To cancel the QCD contribution from NLO PDFs to the NLO EW correction $\Delta \sigma_{EW}$, we normalize the NLO EW relative correction to $\sigma_0$. In this normalization, the NLO EW relative correction $\delta_{EW}$ is practically independent of the PDF set.

\par
Ideally, the NLO QCD+EW correction should be calculated by applying the multi-jet merging approach, which would imply the calculation of NLO QCD and EW corrections to $W^{\pm}ZZ+jet$ final states and thus is beyond the present scope. Therefore, in this paper we calculate the combined NLO QCD+EW correction by using the naive product \cite{Denner-delta}
\begin{eqnarray}
\label{Xection-nlo}
\sigma_{NLO} &=& \sigma_{LO} \left( 1 + \delta_{NLO} \right) \nonumber \\
&=& \sigma_{LO} \left( 1 + \delta_{QCD} \right) \left( 1 + \delta_{EW} \right),
\end{eqnarray}
where $\sigma_{NLO}$ is the NLO QCD+EW corrected cross section and $\delta_{NLO}$ is the NLO QCD+EW relative correction.

\par
In order to keep the convergence of the perturbative QCD description of the $pp \rightarrow W^{\pm} ZZ + X$ processes, we may impose a tight jet veto which can heavily suppress the large QCD correction. We call the event selection scheme with jet veto condition of $p_{T,jet} < p_{T,jet}^{cut} = 50~{\rm GeV}$ as the exclusive scheme (scheme-II) and that without any jet veto as the inclusive scheme (scheme-I). In Table \ref{tab2} and Table \ref{tab3}, we present the LO and NLO QCD+EW corrected cross sections for $pp \rightarrow W^{\pm}ZZ + X$ in both the inclusive and exclusive event selection schemes at the $13$ and $14~{\rm TeV}$ LHC, respectively \footnote{In Table \ref{tab2} and Table \ref{tab3}, $\Delta \sigma_{QCD}$ should be understood as the full NLO QCD correction, i.e., $\Delta \sigma_{QCD} = \sigma_{LO} \delta_{QCD}$.}. The EW corrections from the quark-antiquark and photon-induced channels are shown separately, since they can (in principle) be distinguished by their final states.

\par
From Table \ref{tab2} and Table \ref{tab3} we can see that the photon-induced channels have surprisingly large impact on the NLO EW correction. The NLO EW correction can be heavily suppressed by applying a jet veto. For example, the photon-induced EW relative correction $\delta_{EW}^{q\gamma}$ is $16.23\%$ in the inclusive event collection scheme, but is reduced to $1.73\%$ after applying $p_{T,jet} < 50~{\rm GeV}$ on the final jet, for the $pp \rightarrow W^+ZZ + X$ process at the $13~{\rm TeV}$ LHC. It implies that the theoretical uncertainty from the photon PDF can be reduced by adopting the jet-veto event selection scheme. In our calculation we find that the real light-quark emission correction is the largest NLO QCD contribution and amounts to $56.3\%$ of the full NLO QCD correction for the $W^+ZZ$ production at the $14~ {\rm TeV}$ LHC. In both two event collection schemes, the quark-antiquark and photon-induced EW corrections are negative and positive, respectively. The full NLO EW relative correction to the $W^+ZZ$ production in the inclusive event collection scheme at the $14~ {\rm TeV}$ LHC can reach about $9.67\%$.
\begin{table}[htbp]
\begin{center}
\small
\begin{tabular}{c|cc|cc}
\hline
\multirow{2}{*}{process} & \multicolumn{2}{c|}{$pp \rightarrow W^+ZZ + X$} & \multicolumn{2}{c}{$pp \rightarrow W^-ZZ + X$} \\
\cline{2-5}
& ~scheme-I~ & ~scheme-II~ & ~scheme-I~ & ~scheme-II~ \\
\hline
\hline
$\sigma_{LO}~~$ $[fb]$ & 17.078(5) & 17.078(5) & 8.625(2) & 8.625(2) \\
$\sigma_{NLO}$ $[fb]$ & 42.71(5) & 20.68(3) & 24.19(2) & 11.39(2) \\
\hline
~$\Delta \sigma_{QCD}$ $[fb]$~ & 22.06(4) & 4.79(3) & 12.82(2) & 3.29(2) \\
$\Delta \sigma_{EW}^{q\bar{q}}~$ $[fb]$ & -1.412(6) & -1.412(6) & -0.710(3) & -0.710(3) \\
$\Delta \sigma_{EW}^{q\gamma}~$ $[fb]$ & 3.222(2) & 0.3443(4) & 2.061(1) & 0.2335(3) \\
\hline
$\delta_{QCD}$ $[\%]$ & 129.2 & 28.0 & 148.6 & 38.1 \\
$\delta_{EW}^{q\bar{q}}~$ $[\%]$ & -7.11 & -7.11 & -6.74 & -6.74 \\
$\delta_{EW}^{q\gamma}~$ $[\%]$ & 16.23 & 1.73 & 19.57 & 2.22 \\
$\delta_{NLO}$ $[\%]$ & 150.1 & 21.1 & 180.5 & 32.0 \\
\hline
\end{tabular}
 \caption{\small \label{tab2}{The LO, NLO QCD+EW corrected integrated cross sections for $pp \rightarrow W^{\pm}ZZ + X$ and the corresponding NLO corrections at the $\sqrt{S}= 13~{\rm TeV}$ LHC.}}
\end{center}
\end{table}
\begin{table}[htbp]
\begin{center}
\small
\begin{tabular}{c|cc|cc}
\hline
\multirow{2}{*}{process} & \multicolumn{2}{c|}{$pp \rightarrow W^+ZZ + X$} & \multicolumn{2}{c}{$pp \rightarrow W^-ZZ + X$} \\
\cline{2-5}
& ~scheme-I~ & ~scheme-II~ & ~scheme-I~ & ~scheme-II~ \\
\hline
\hline
$\sigma_{LO}~~$ $[fb]$ & 19.126(5) & 19.126(5) & 9.816(3) & 9.816(3) \\
$\sigma_{NLO}$ $[fb]$ & 48.96(4) & 23.05(3) & 28.19(2) & 12.87(2) \\
\hline
~$\Delta \sigma_{QCD}$ $[fb]$~ & 25.52(4) & 5.25(3) & 15.05(2) & 3.67(2) \\
$\Delta \sigma_{EW}^{q\bar{q}}~$ $[fb]$ & -1.601(6) & -1.601(6) & -0.819(3) & -0.819(3) \\
$\Delta \sigma_{EW}^{q\gamma}~$ $[fb]$ & 3.753(2) & 0.3952(5) & 2.424(1) & 0.2702(3) \\
\hline
$\delta_{QCD}$ $[\%]$ & 133.4 & 27.4 & 153.3 & 37.4 \\
$\delta_{EW}^{q\bar{q}}~$ $[\%]$ & -7.20 & -7.20 & -6.83 & -6.83 \\
$\delta_{EW}^{q\gamma}~$ $[\%]$ & 16.87 & 1.78 & 20.22 & 2.25 \\
$\delta_{NLO}$ $[\%]$ & 156.0 & 20.5 & 187.2 & 31.1 \\
\hline
\end{tabular}
 \caption{\small \label{tab3}{The LO, NLO QCD+EW corrected integrated cross sections for $pp \rightarrow W^{\pm}ZZ + X$ and the corresponding NLO corrections at the $\sqrt{S}= 14~{\rm TeV}$ LHC.}}
\end{center}
\end{table}

\par
\begin{figure}[htbp]
  \begin{center}
      \includegraphics[scale=0.23]{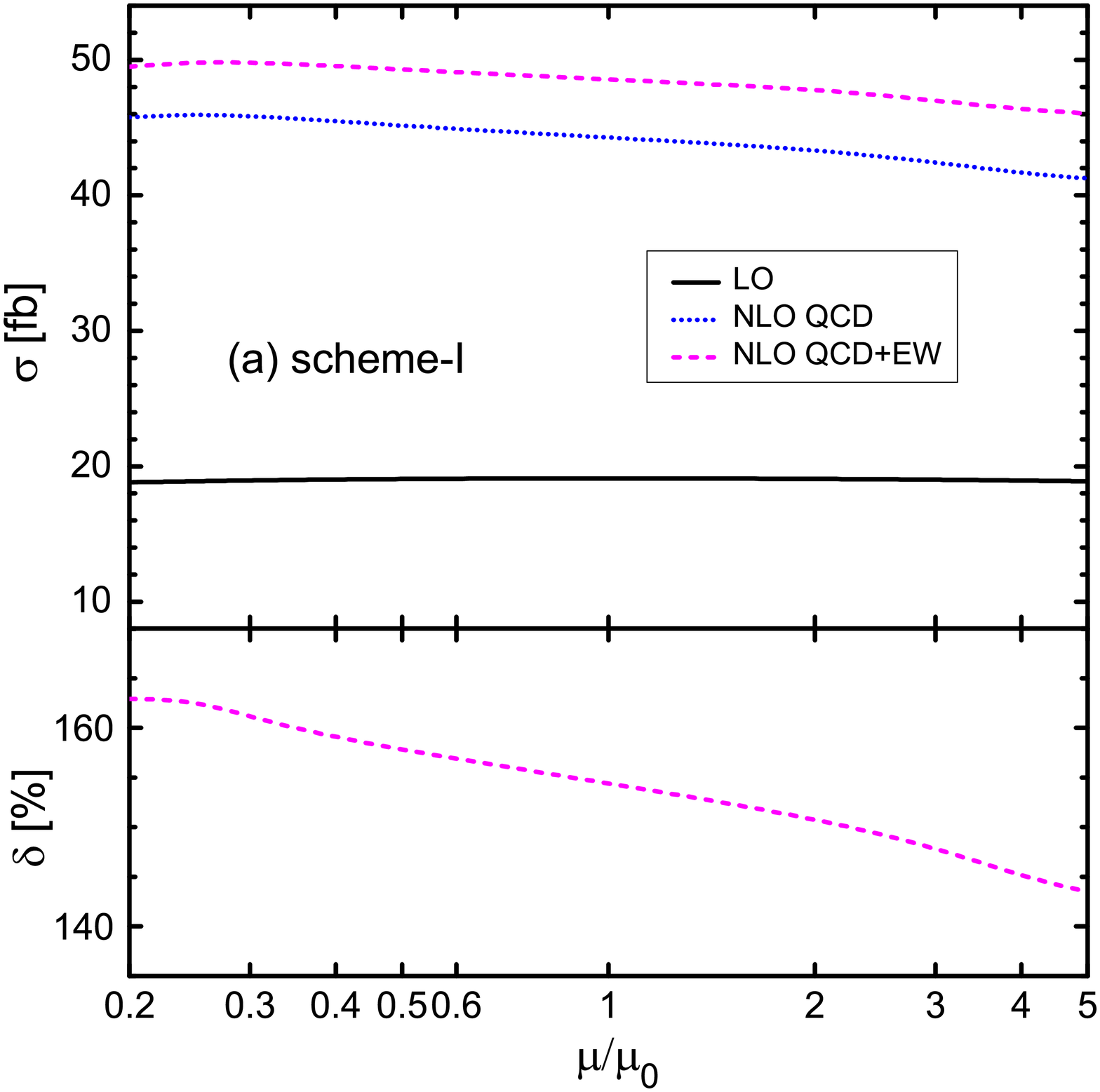}
      \includegraphics[scale=0.23]{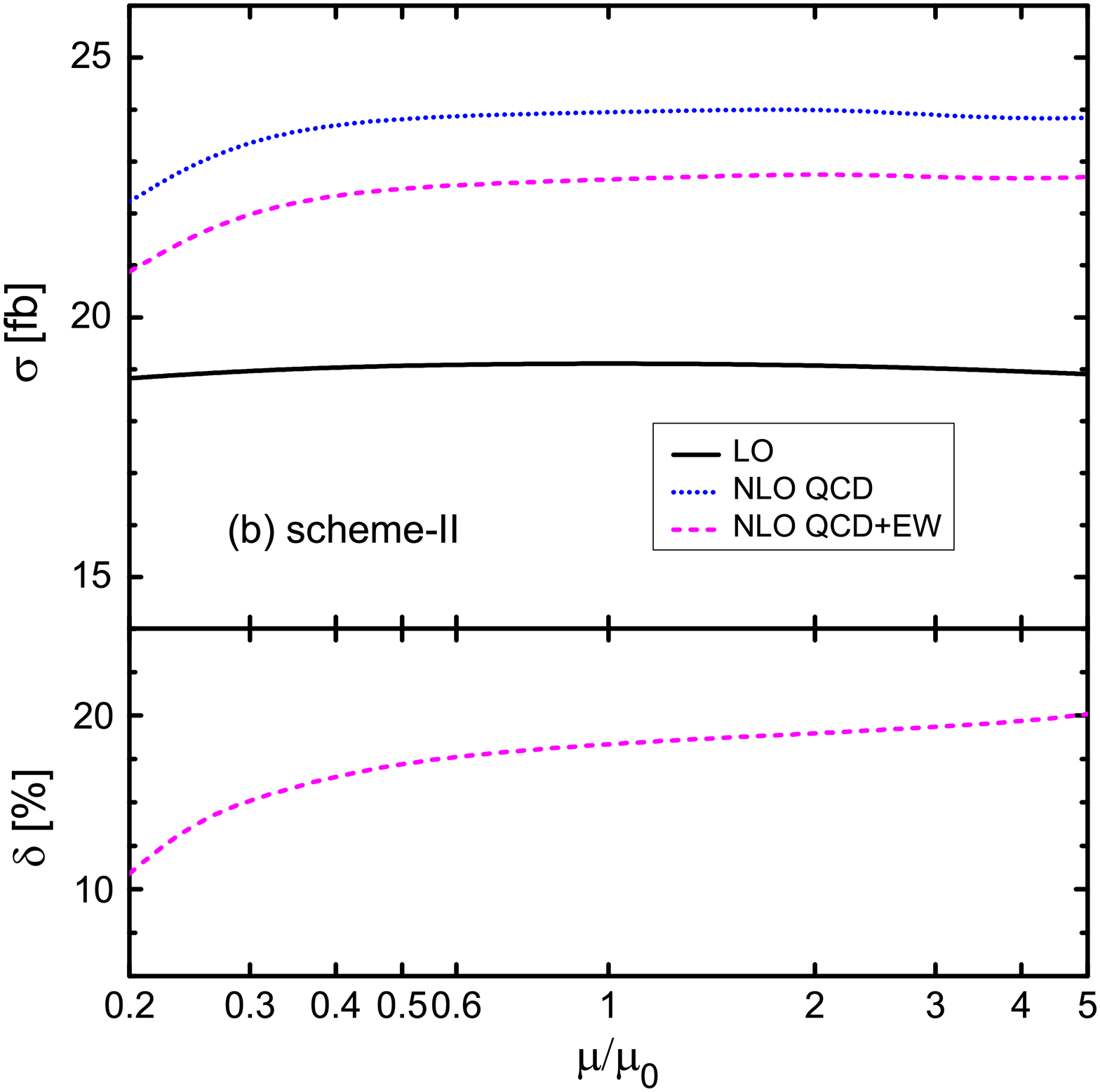}
    \caption{\small The fractorization/renormalization scale dependence of $\sigma_{LO}$ (solid), $\sigma_{QCD}$ (dotted) and $\sigma_{NLO}$ (dashed) for the $W^+ZZ$ production at the $\sqrt{S}= 14~{\rm TeV}$ LHC in the (a) inclusive and (b) exclusive event selection schemes. }
   \label{fig:scale}
  \end{center}
\end{figure}

Now we turn to the fractorization/renormalization scale dependence of the LO and NLO corrected integrated cross sections. The factorization scale $\mu_F$ affects both the LO and NLO corrected cross sections via the factorization procedure due to the $\mu_F$ dependence of the parton densities. The renormalization scale $\mu_R$ occurs in higher order perturbative calculation via the renormalization procedure and strongly affects the QCD correction. For simplicity, we set $\mu_F = \mu_R = \mu$ in our calculation. In Figs.\ref{fig:scale}(a) and (b), we depict the LO, NLO QCD and QCD+EW corrected cross sections ($\sigma_{LO}$, $\sigma_{QCD}$ and $\sigma_{NLO}$) for the $W^+ZZ$ production at the $14~ {\rm TeV}$ LHC as functions of $\mu$ by adopting the inclusive and exclusive event selection schemes, respectively. The corresponding NLO QCD+EW relative corrections are shown in the lower plots. To estimate the theoretical uncertainty from the factorization/renormalization scale quantitatively, we introduce the scale uncertainty as
\begin{eqnarray}
\eta = \frac{\max\left\{\sigma(\mu) | 0.2 \mu_0 \leq \mu \leq 5 \mu_0 \right\} - \min\left\{\sigma(\mu) | 0.2 \mu_0 \leq \mu \leq 5 \mu_0 \right\}}{\sigma(\mu_0)}.
\end{eqnarray}
From Figs.\ref{fig:scale} (a,b) we obtain $\eta = 1.6\%$ at the LO, and $\eta = 10.8\%$ and $7.5\%$ at the QCD NLO in the inclusive and exclusive event collection schemes respectively. We can see that the scale uncertainty at the LO is much less than at the QCD NLO, because the strong coupling $\alpha_s$ is not involved in the LO matrix elements. When we include both the NLO QCD and NLO EW corrections into consideration, the scale uncertainties are $7.9\%$ and $8.4\%$ in the inclusive and exclusive event selection schemes, respectively. We can conclude that the scale uncertainty at the QCD+EW NLO mainly comes from the QCD correction.

\par
From the numerical results given in this subsection, we can draw the following four conclusions:
(1) The NLO QCD correction is very large and ruins the convergence of the perturbative QCD description in the inclusive event selection scheme.
(2) The NLO QCD correction mainly comes from the real jet radiation subprocesses, and therefore can be heavily reduced by adopting the exclusive event selection scheme with a jet veto. But a tight jet veto would introduce a new source of theoretical uncertainties from various other processes.
(3) Due to the small scale dependence of PDFs in the Feynman-$x$ region, the scale dependence of the LO cross section is not apparent. But the LO scale uncertainty does not give a good estimate of the higher order QCD contribution.
(4) Compared to the NLO QCD correction, the NLO EW correction is insensitive to the factorization/renormalization scale.

\par
\subsection{Kinematic distributions}
\par
In this subsection we present some kinematic distributions of final produced particles for the $W^+ZZ$ production at the $14~ {\rm TeV}$ LHC. In order to take into account the off-shell contributions and spin correlations from the $W^+$- and $Z$-boson leptonic decays, we transform the differential cross sections into Les Houches event files \cite{lhe} and use MadSpin \cite{madspin} to obtain events after the vector boson decays. For each kinematic variable $x$ considered in the following, we provide the LO, NLO QCD and QCD+EW corrected distributions, i.e., $d\sigma_{LO}/dx$, $d\sigma_{QCD}/dx$ and $d\sigma_{NLO}/dx$.

\par
\begin{figure}[htbp]
  \begin{center}
      \includegraphics[scale=0.25]{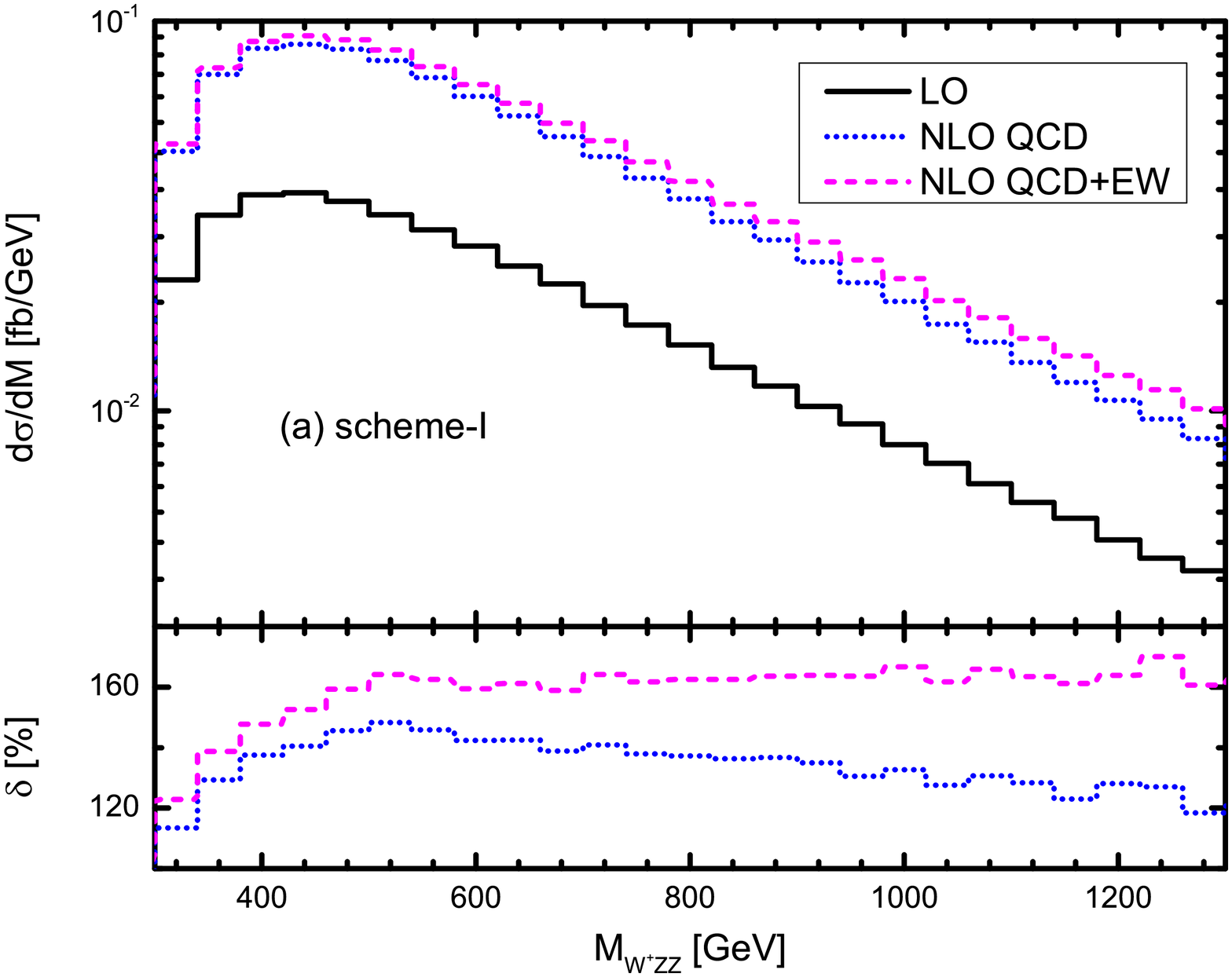}
      \includegraphics[scale=0.25]{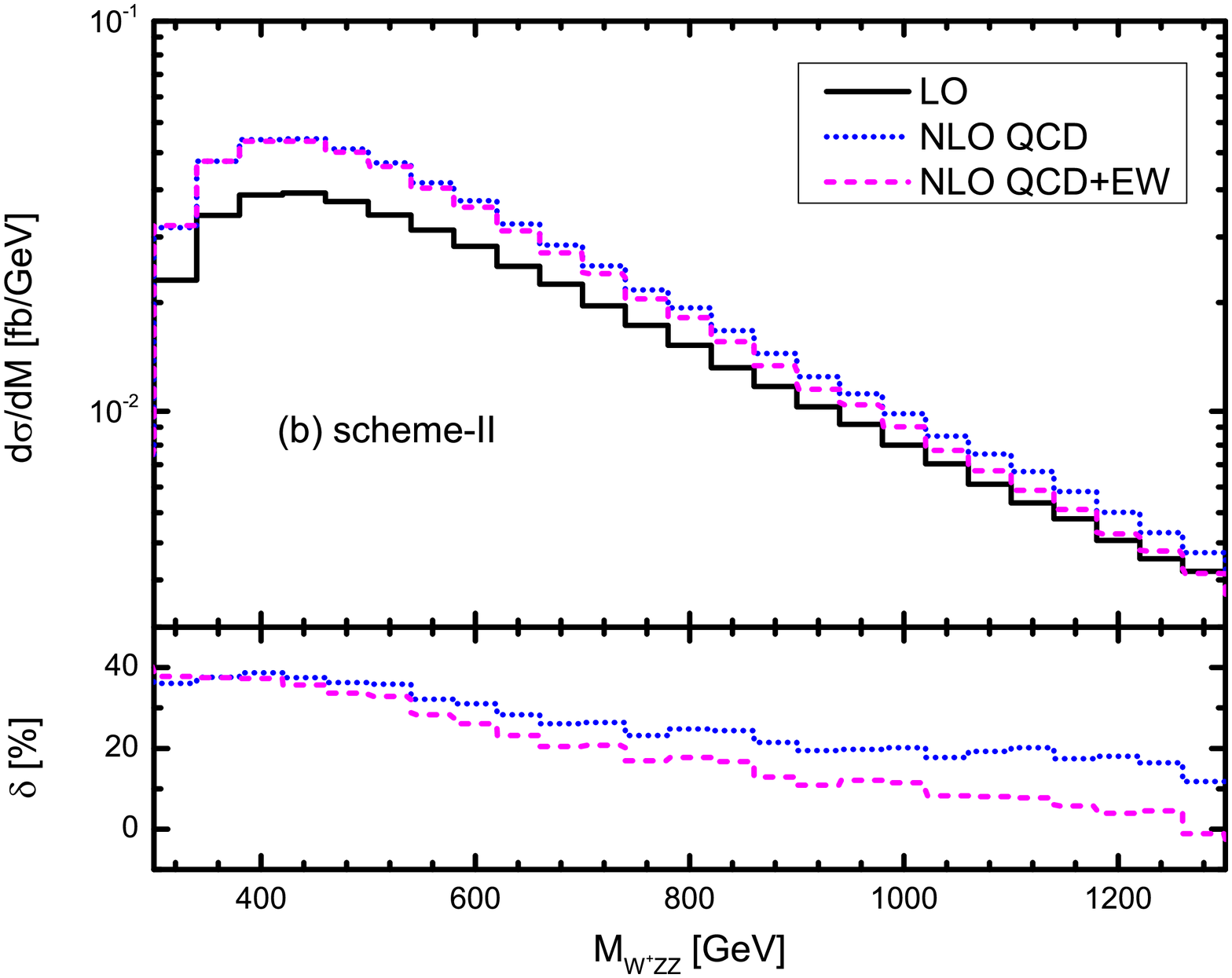}
   \caption{\small The $W^+ZZ$ invariant mass distributions $d\sigma_{LO}/dM_{W^+ZZ}$ (solid), $d\sigma_{QCD}/dM_{W^+ZZ}$ (dotted), $d\sigma_{NLO}/dM_{W^+ZZ}$ (dashed) and the corresponding relative corrections for $pp \rightarrow W^+ZZ + X$ at the $\sqrt{S} = 14~{\rm TeV}$ LHC in the (a) inclusive and (b) exclusive event selection schemes. }
   \label{fig:m_wzz}
  \end{center}
\end{figure}

We present the invariant mass distributions of $W^+ZZ$ system by adopting the inclusive and exclusive event collection schemes in Figs.\ref{fig:m_wzz}(a) and (b), respectively. The corresponding relative corrections are depicted in the nether plots. The figures show that the $W^+ZZ$ invariant mass distributions in both event selection schemes reach their maxima at the position of $M_{W^+ZZ} \sim 440~{\rm GeV}$, and the inclusive NLO QCD correction enhances the LO $M_{W^+ZZ}$ distribution significantly. From Fig.\ref{fig:m_wzz}(a) we see that the NLO EW correction in the inclusive event selection scheme is positive but very small compared with the corresponding NLO QCD correction in the region of $M_{W^+ZZ} > 400~ {\rm GeV}$. From Fig.\ref{fig:m_wzz}(b) we find that in the exclusive event selection scheme the NLO QCD correction is heavily reduced and the NLO EW correction becomes negative because of the jet veto in the exclusive event selection scheme. In the plotted $M_{W^+ZZ}$ region, the NLO QCD+EW relative corrections in the inclusive and exclusive event selection schemes range from $123\%$ to $170\%$ and from $-1\%$ to $38\%$, respectively. The NLO QCD+EW relative correction to the $M_{W^+ZZ}$ distribution in the inclusive event selection scheme is more dependent on the phase space than in the exclusive event selection scheme, because the NLO QCD correction from the gluon-induced channels is the dominant contribution at the NLO and is more closely related to the phase space. We conclude that the real jet radiation would induce large NLO contribution in the inclusive event collection scheme, and we can keep the convergence of the perturbative QCD description and obtain moderate NLO QCD+EW correction by adopting the exclusive event selection scheme. However, a jet veto as tight as $p_{T,jet} < 50~{\rm GeV}$ introduces a new source of theoretical uncertainties from various processes. Such uncertainties could be analyzed by dedicated jet veto resummations.

\par
\begin{figure}[htbp]
  \begin{center}
      \includegraphics[scale=0.25]{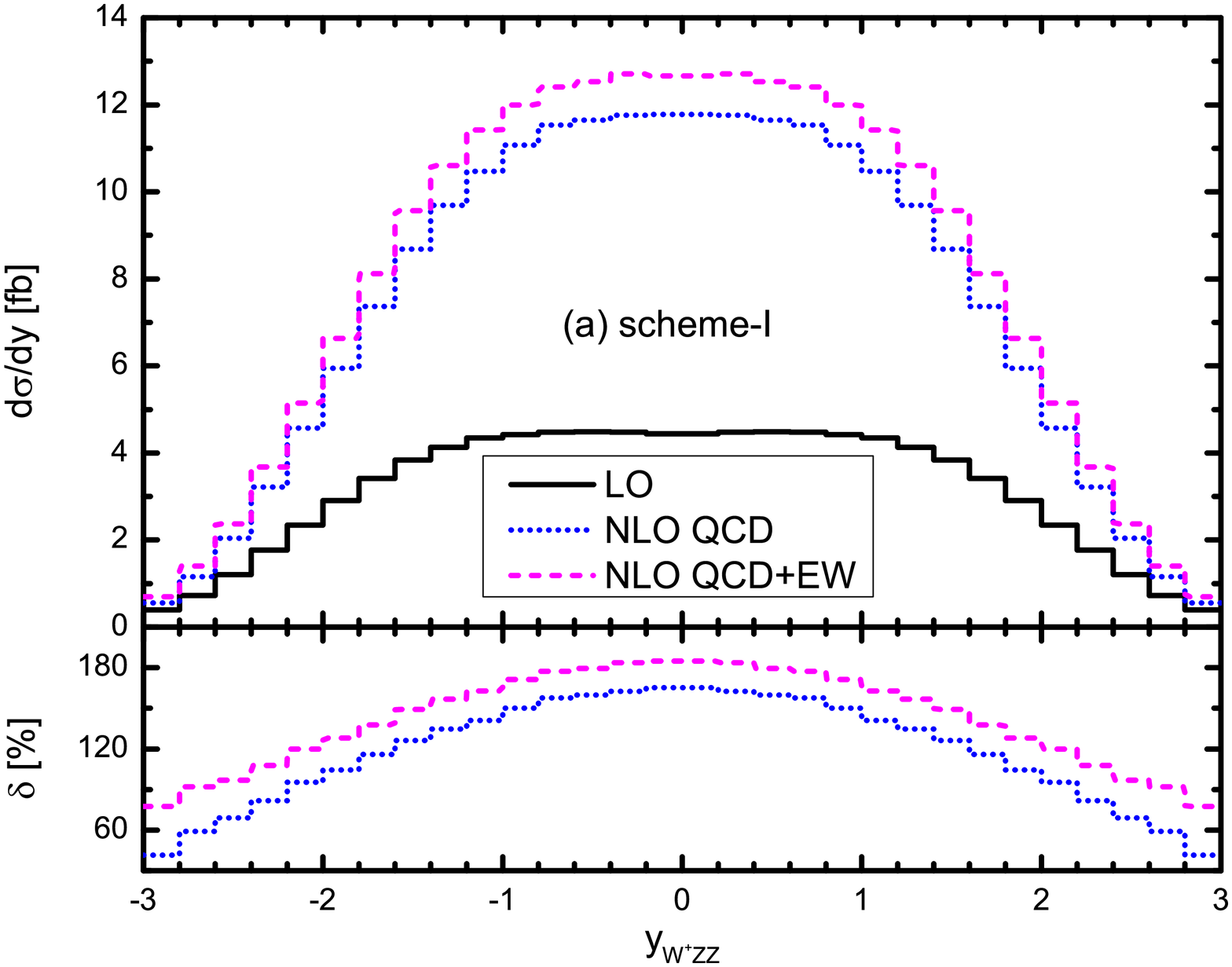}
      \includegraphics[scale=0.25]{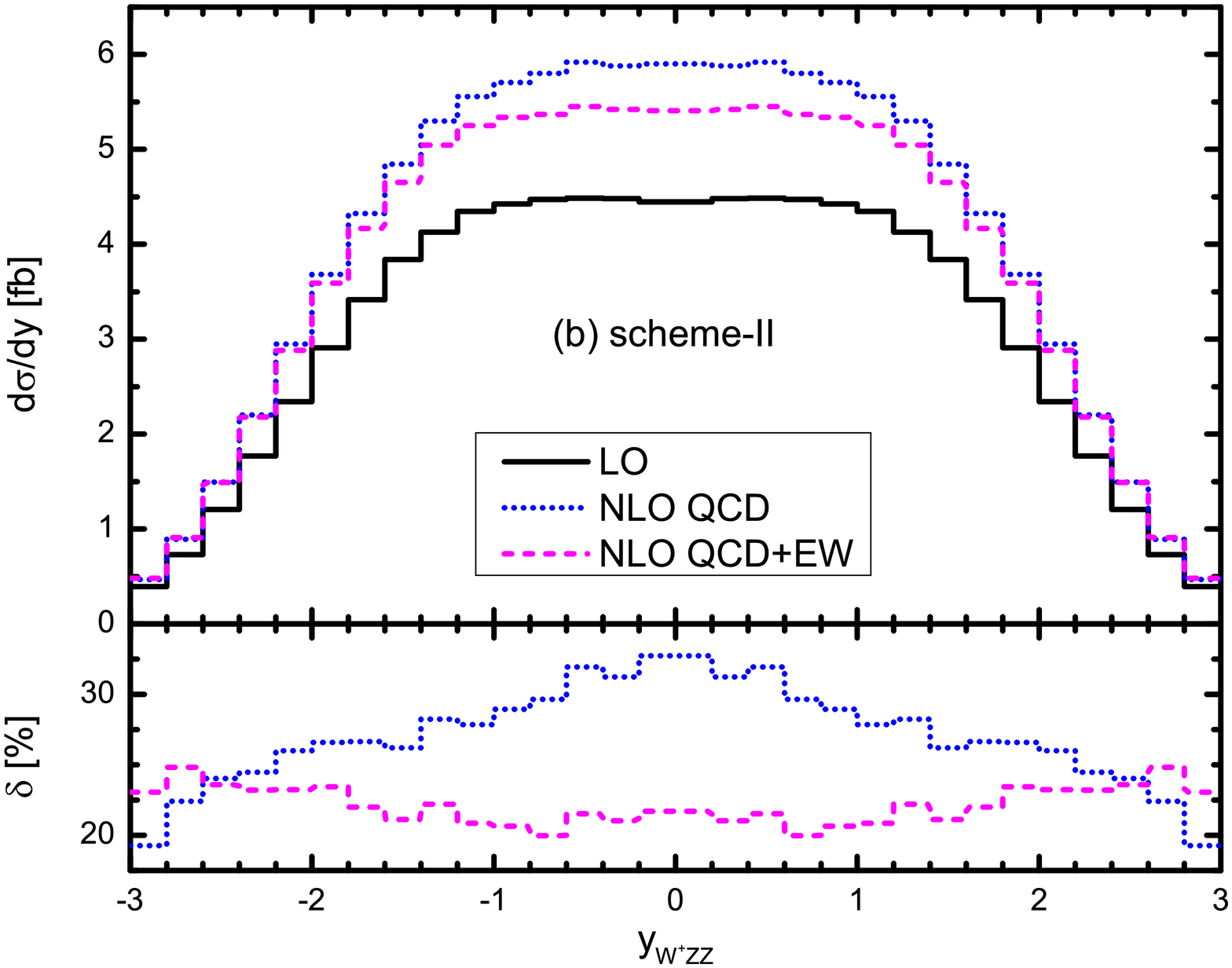}
    \caption{\small
    The $W^+ZZ$ rapidity distributions $d\sigma_{LO}/dy_{W^+ZZ}$ (solid), $d\sigma_{QCD}/dy_{W^+ZZ}$ (dotted), $d\sigma_{NLO}/dy_{W^+ZZ}$ (dashed) and the corresponding relative corrections for $pp \rightarrow W^+ZZ + X$ at the $\sqrt{S} = 14~{\rm TeV}$ LHC in the (a) inclusive and (b) exclusive event selection schemes.  }
   \label{fig:y wzz}
  \end{center}
\end{figure}

The rapidity distributions of $W^+ZZ$ system by adopting the inclusive and exclusive event collection schemes are shown in Figs.\ref{fig:y wzz}(a) and (b) separately. The corresponding relative corrections are plotted in the lower panels. From the figures we can see that the NLO QCD relative corrections in the inclusive and exclusive event collection schemes are about $165\%$ and $33\%$, respectively, at the position of $y_{W^+ZZ} = 0$. The NLO EW correction enhances the LO $W^+ZZ$ rapidity distribution a little bit in the inclusive event selection scheme, but reduces the LO $y_{W^+ZZ}$ distribution in the exclusive event selection scheme. In the inclusive event selection scheme the NLO QCD correction is much larger than the NLO EW correction. However, the NLO QCD correction in the exclusive event collection scheme is heavily reduced, particularly in the region of $|y_{W^+ZZ}| < 1$. As we know, the events with large $W^+ZZ$ transverse momentum, i.e., $p_{T,W^+ZZ} > 50~{\rm GeV}$, tend to be produced more centrally, i.e., $y_{W^+ZZ} \rightarrow 0$, and will be excluded in the exclusive event selection scheme due to the jet veto and the conservation of transverse momentum. That's why the NLO QCD correction in the exclusive event selection scheme is small in the range of $|y_{W^+ZZ}| < 1$. The NLO QCD+EW correction in the exclusive event selection scheme suppresses the LO $y_{W^+ZZ}$ distribution in the vicinity of $y_{W^+ZZ} = 0$.

\par
\begin{figure}[htbp]
  \begin{center}
      \includegraphics[scale=0.25]{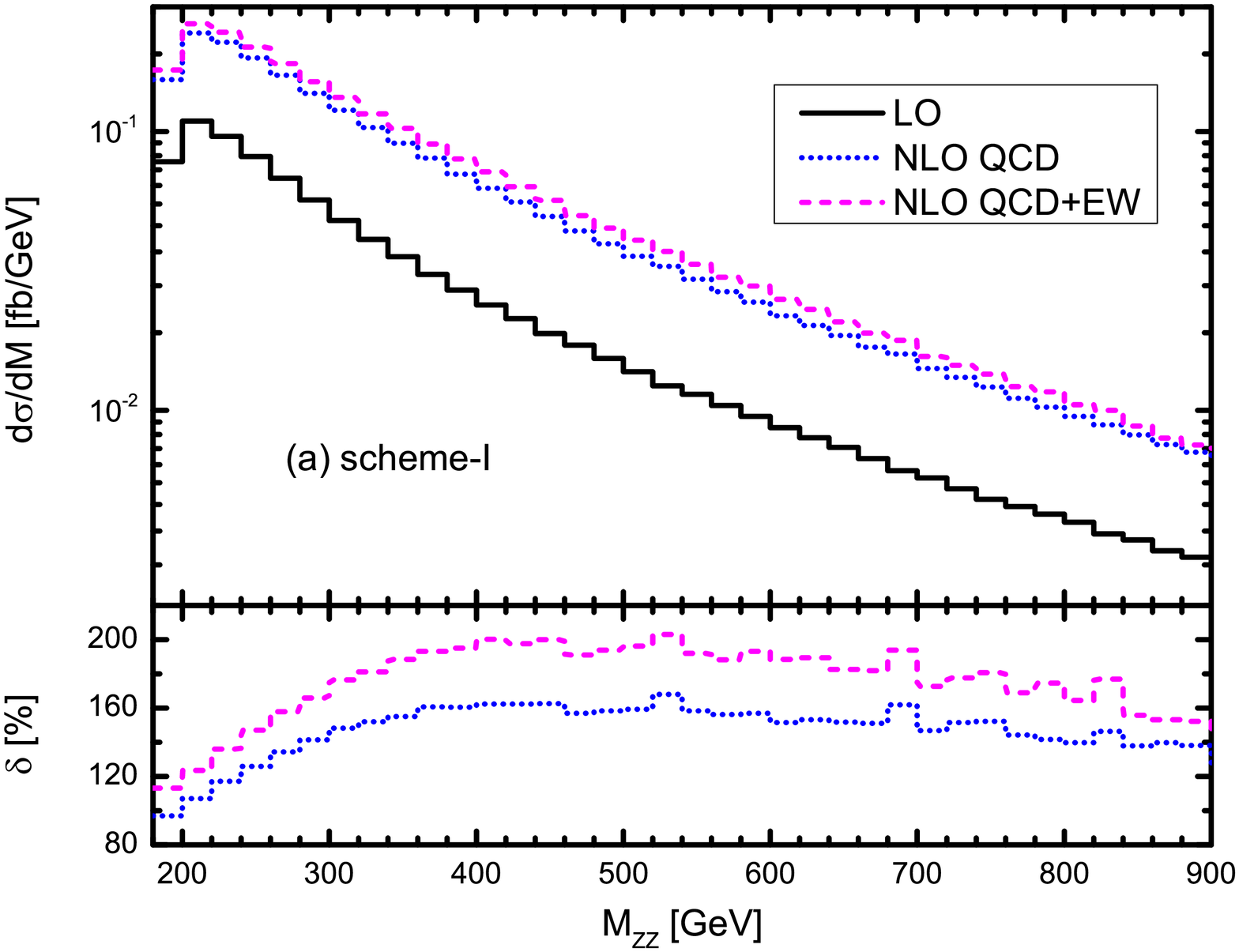}
      \includegraphics[scale=0.25]{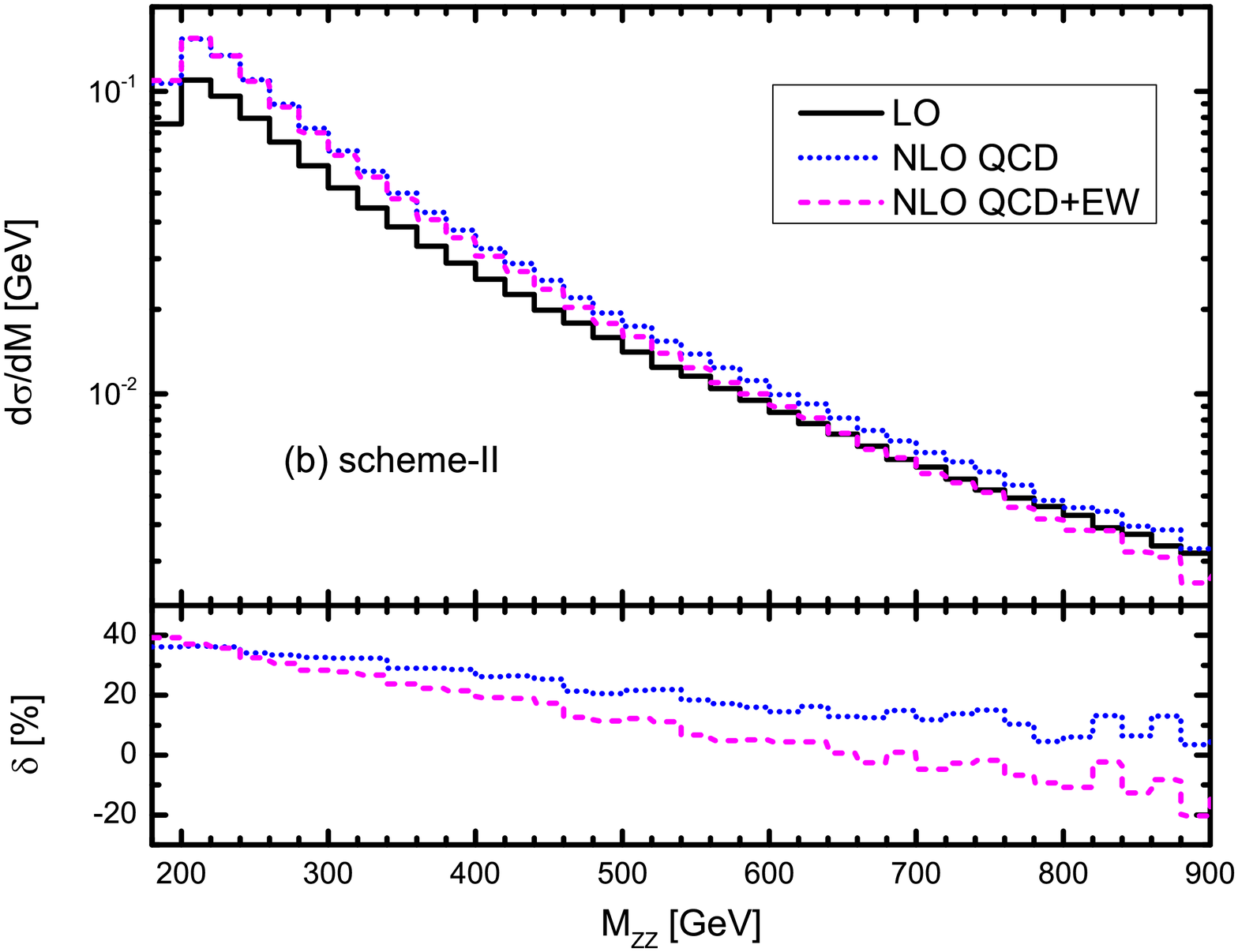}
    \caption{\small The $Z$-pair invariant mass distributions $d\sigma_{LO}/dM_{ZZ}$ (solid), $d\sigma_{QCD}/dM_{ZZ}$ (dotted), $d\sigma_{NLO}/dM_{ZZ}$ (dashed) and the corresponding relative corrections for $pp \rightarrow W^+ZZ + X$ at the $\sqrt{S} = 14~{\rm TeV}$ LHC in the (a) inclusive and (b) exclusive event selection schemes.     }
   \label{fig:m_zz}
  \end{center}
\end{figure}

The $Z$-pair invariant mass distributions by adopting the inclusive and exclusive event collection criteria are plotted in Figs.\ref{fig:m_zz}(a) and (b), respectively. Form the figures we see that the NLO QCD and EW corrections do not distort the line shape of the LO $M_{ZZ}$ distribution, and both the LO and NLO corrected $M_{ZZ}$ distributions reach their maxima at $M_{ZZ} \sim 210~ {\rm GeV}$. The NLO EW correction slightly enhances and suppresses the LO $M_{ZZ}$ distribution in the inclusive and exclusive event collection schemes, respectively, in the plotted $M_{ZZ}$ region. The NLO QCD relative correction can exceed $138\%$ when $M_{ZZ} > 300~{\rm GeV}$ in the inclusive event collection scheme, and is less than $36\%$ in the whole plotted $M_{ZZ}$ region in the exclusive event collection scheme. It indicates that the NLO QCD correction to the $M_{ZZ}$ distribution in the inclusive event collection scheme could be very large and destroy the perturbative description, and the perturbative convergence can be improved by applying a tight jet veto in the exclusive event collection scheme.

\par
\begin{figure}[htbp]
  \begin{center}
      \includegraphics[scale=0.25]{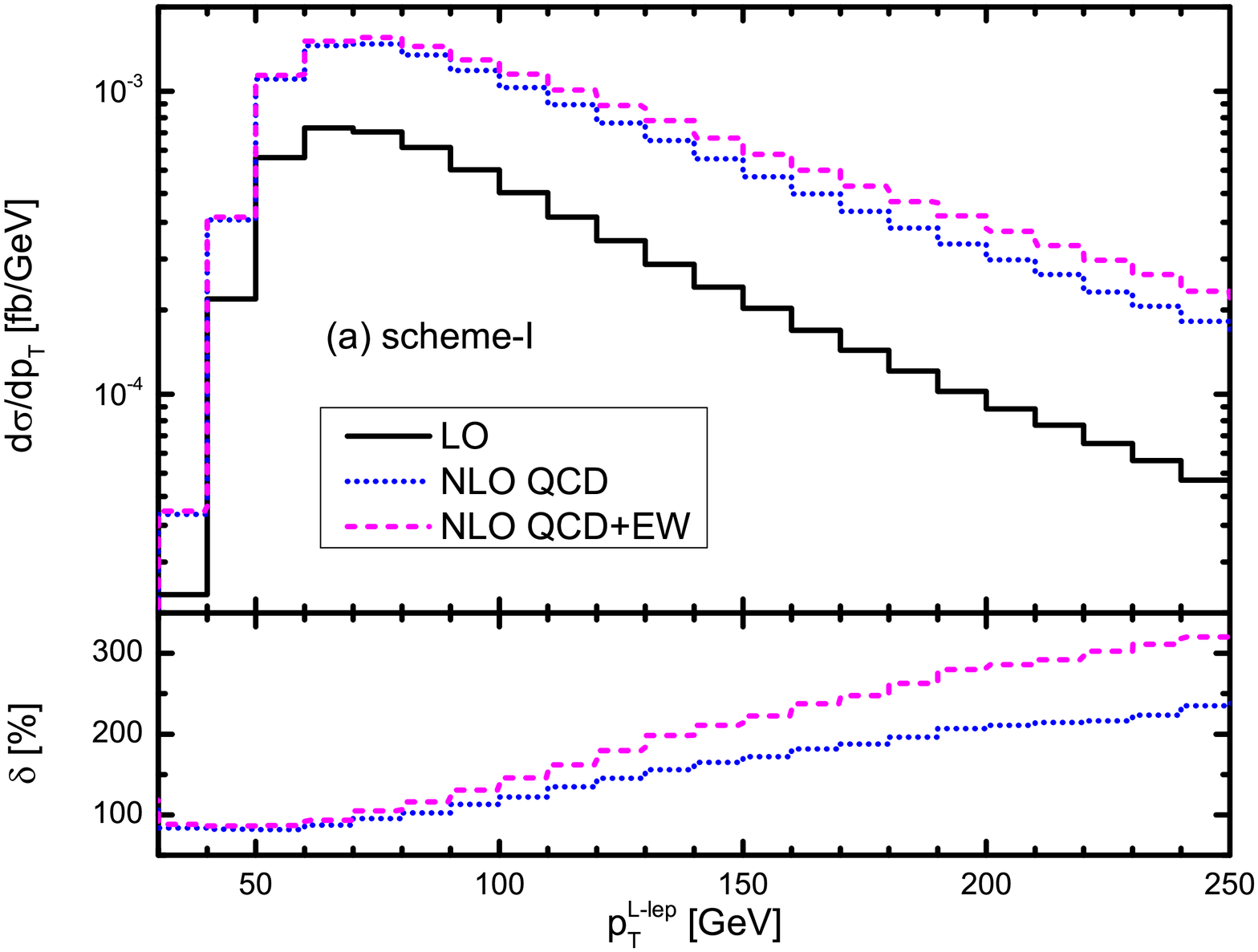}
      \includegraphics[scale=0.25]{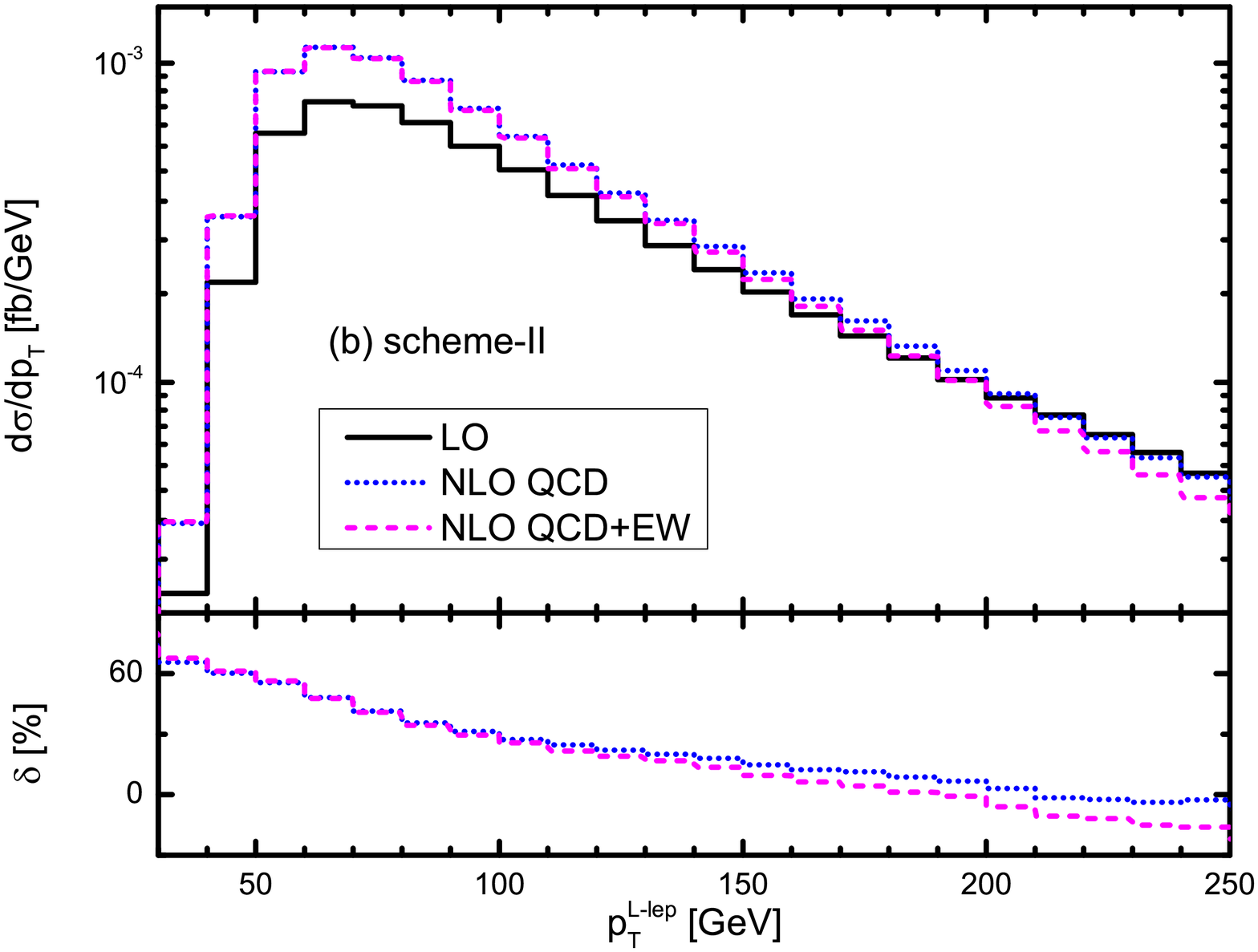}
    \caption{\small The leading lepton transverse momentum distributions $d\sigma_{LO}/dp_T^{L-lep}$ (solid), $d\sigma_{QCD}/dp_T^{L-lep}$ (dotted), $d\sigma_{NLO}/dp_T^{L-lep}$ (dashed) and the corresponding relative corrections for $pp \rightarrow W^+ZZ \rightarrow \ell^+_1 \nu_{\ell_1} \ell^+_2 \ell^-_2 \ell^+_3 \ell^-_3 + X$ at the $\sqrt{S} = 14~{\rm TeV}$ LHC in the (a) inclusive and (b) exclusive event selection schemes. }
   \label{fig:pt_l}
  \end{center}
\end{figure}
\begin{figure}[htbp]
  \begin{center}
      \includegraphics[scale=0.25]{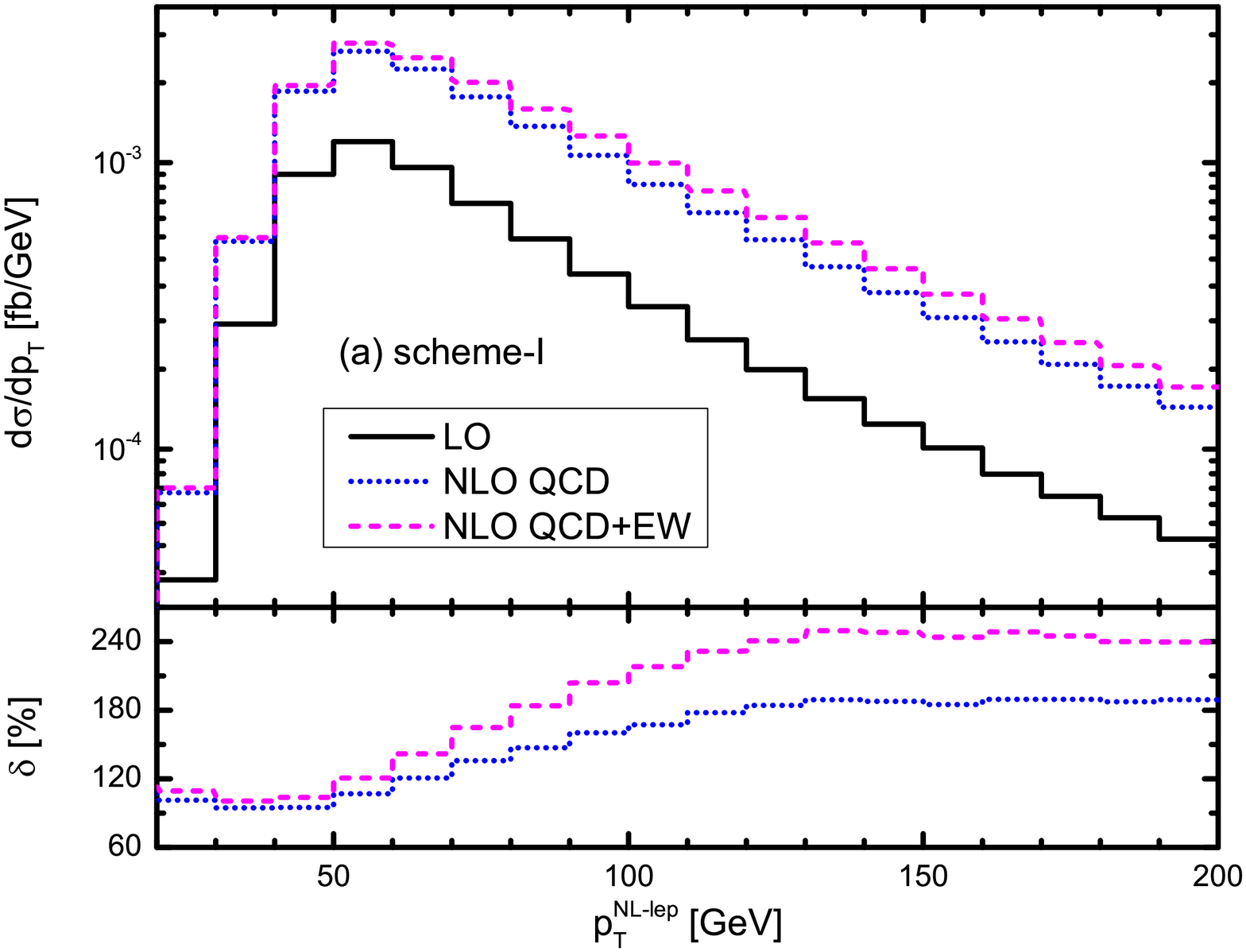}
      \includegraphics[scale=0.25]{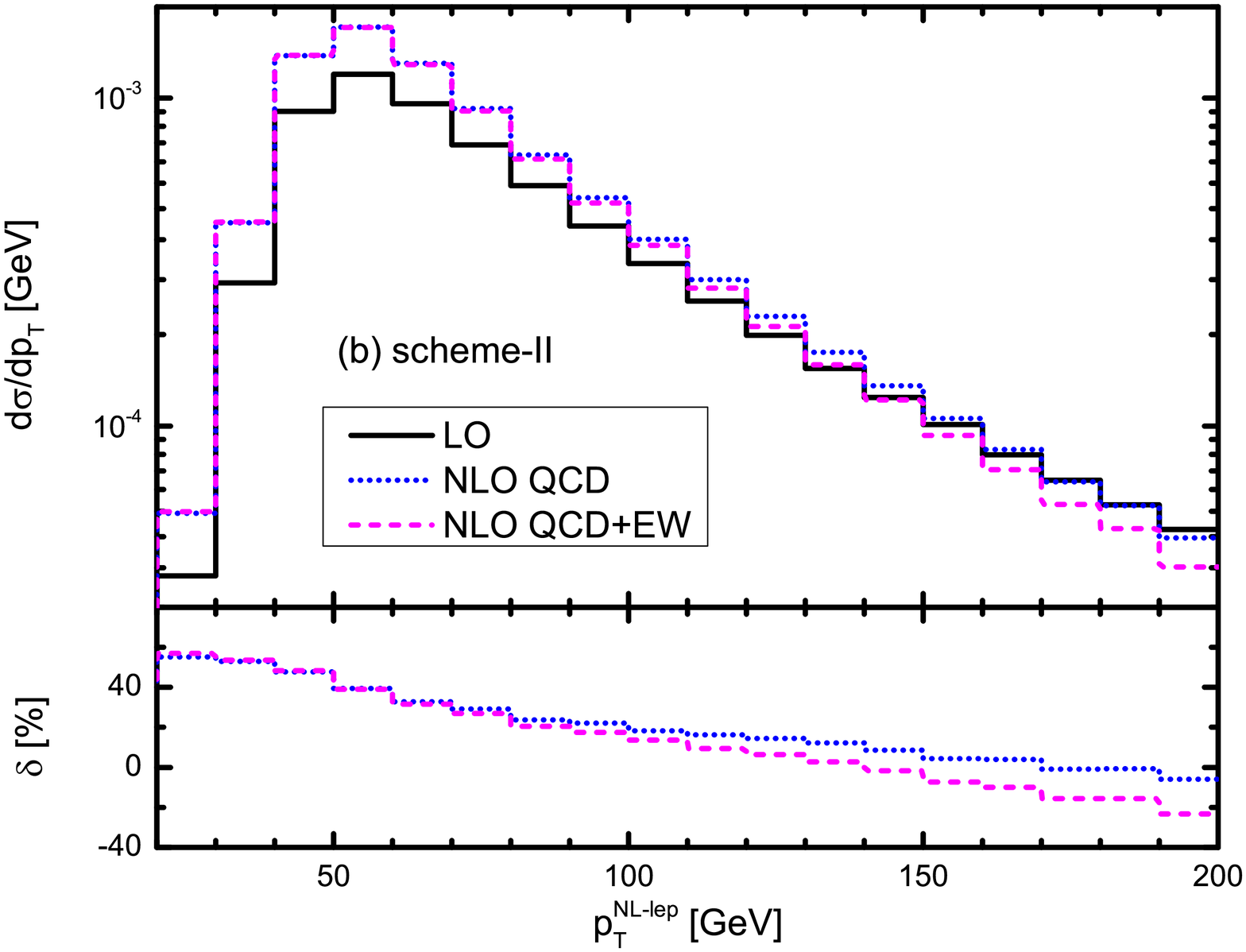}
    \caption{\small The next-to-leading lepton transverse momentum distributions $d\sigma_{LO}/dp_T^{NL-lep}$ (solid), $d\sigma_{QCD}/dp_T^{NL-lep}$ (dotted), $d\sigma_{NLO}/dp_T^{NL-lep}$ (dashed) and the corresponding relative corrections for $pp \rightarrow W^+ZZ \rightarrow \ell^+_1 \nu_{\ell_1} \ell^+_2 \ell^-_2 \ell^+_3 \ell^-_3 + X$ at the $\sqrt{S} = 14~{\rm TeV}$ LHC in the (a) inclusive and (b) exclusive event selection schemes. }
   \label{fig:pt_nl}
  \end{center}
\end{figure}
\begin{figure}[htbp]
  \begin{center}
      \includegraphics[scale=0.25]{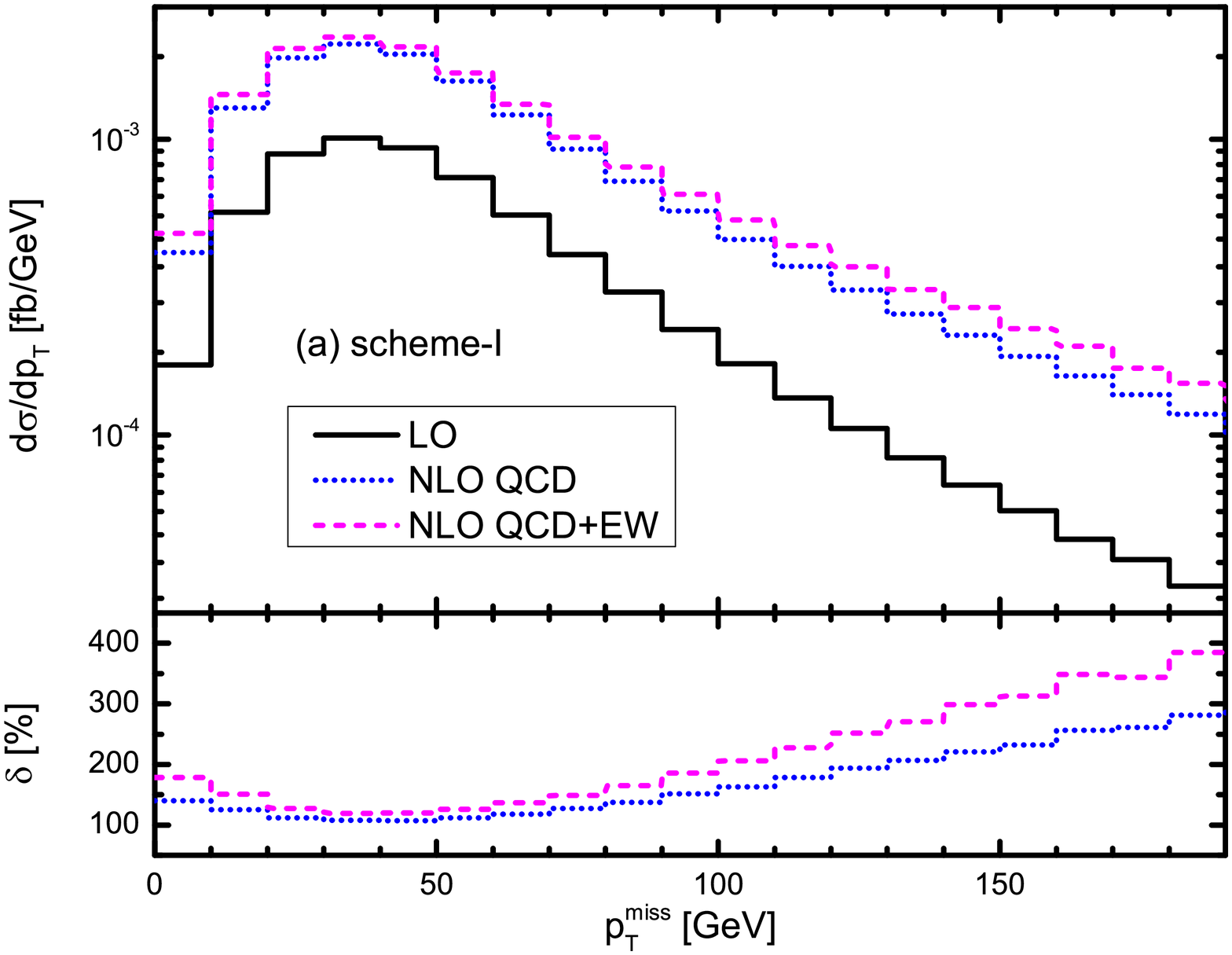}
      \includegraphics[scale=0.25]{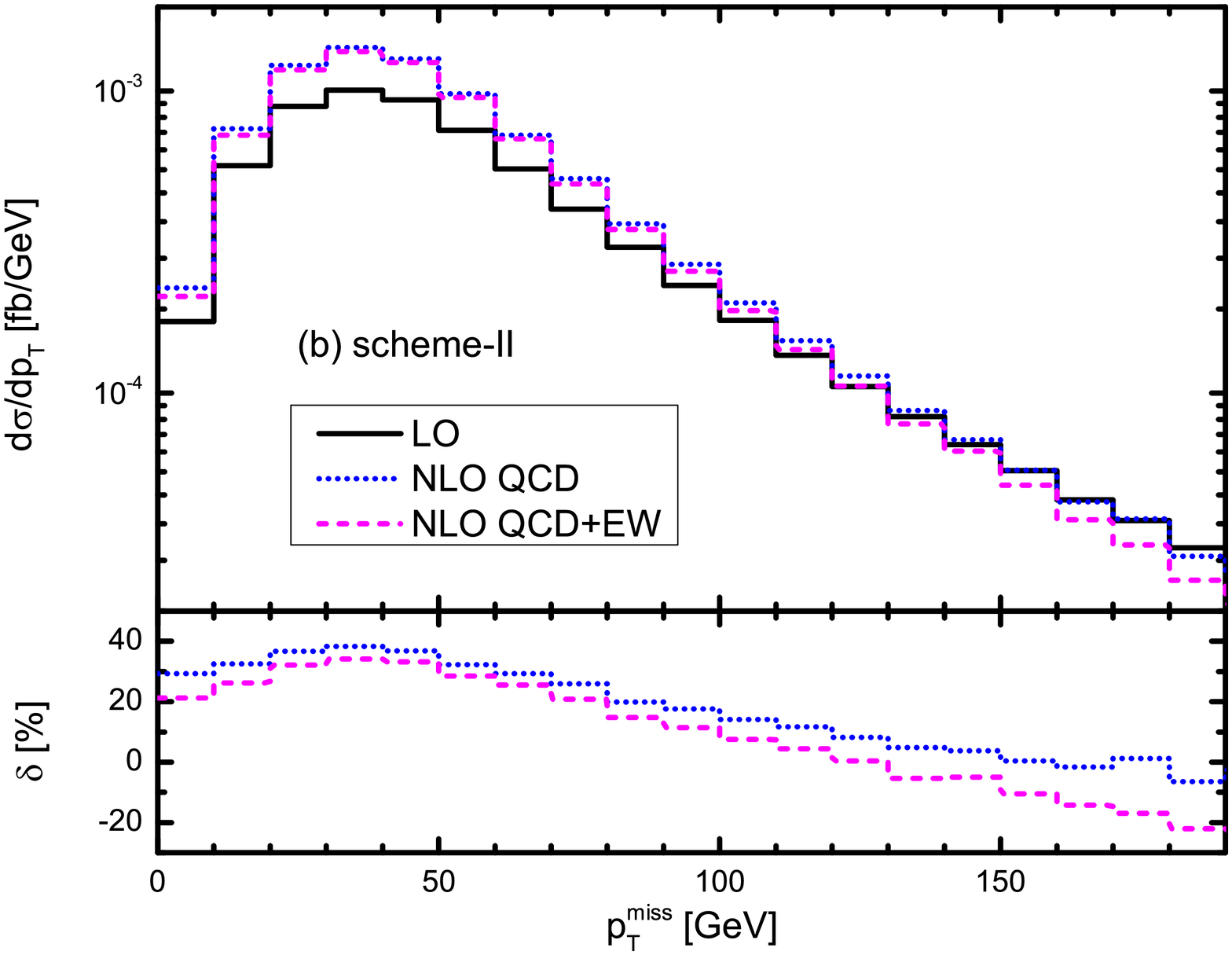}
    \caption{\small The missing transverse momentum distributions $d\sigma_{LO}/dp_T^{miss}$ (solid), $d\sigma_{QCD}/dp_T^{miss}$ (dotted), $d\sigma_{NLO}/dp_T^{miss}$ (dashed) and the corresponding relative corrections for $pp \rightarrow W^+ZZ \rightarrow \ell^+_1 \nu_{\ell_1} \ell^+_2 \ell^-_2 \ell^+_3 \ell^-_3 + X$ at the $\sqrt{S} = 14~{\rm TeV}$ LHC in the (a) inclusive and (b) exclusive event selection schemes.   }
   \label{fig:pt_missing}
  \end{center}
\end{figure}

Concerning the sequential leptonic decays of the final $W^+$ and $Z$-bosons, we study the $pp \rightarrow W^+ZZ \rightarrow \ell^+_1 \nu_{\ell_1} \ell^+_2 \ell^-_2 \ell^+_3 \ell^-_3 + X$ process by adopting the improved narrow width approximation. The branch ratios for the $W$- and $Z$-boson leptonic decay modes are obtained by using the MadSpin program. We define the final lepton with the largest transverse momentum among all leptons as the leading lepton and that with the second largest transverse momentum as the next-to-leading lepton. In Figs.\ref{fig:pt_l}(a,b), Figs.\ref{fig:pt_nl}(a,b) and Figs.\ref{fig:pt_missing}(a,b) we display the distributions of the leading lepton, next-to-leading lepton and missing transverse momenta in the inclusive and exclusive event collection schemes separately. The corresponding relative corrections are drawn in the nether panels. We see from these figures that all the transverse momentum distributions of the leading lepton, next-to-leading lepton and missing energy have similar behavior. Both the NLO QCD and NLO EW corrections enhance the LO $p_T$ distributions in the whole plotted $p_T$ region in the inclusive event selection scheme, but suppress the LO $p_T$ distributions in high $p_T$ region in the exclusive event selection scheme. In the inclusive event selection scheme, the NLO QCD relative correction increases with the increment of $p_T$ in the region of $p_T > 50~{\rm GeV}$. For example, we can read out from Fig.\ref{fig:pt_l}(a) that the NLO QCD relative correction in the inclusive event selection scheme increases from $82\%$ to $235\%$ as the increment of $p_{T}^{L-lep}$ from $50~{\rm GeV}$ to $250~{\rm GeV}$. The large NLO QCD correction in high $p_T$ region in the inclusive event selection scheme is dominated by the gluon-induced channels \cite{WWW-WZZ,EWWWZ}. In the exclusive event selection scheme, the NLO EW correction is very small and negligible compared with the corresponding QCD correction in the whole plotted $p_T$ region. The NLO QCD relative correction increases in low $p_T$ region and then drops down in high $p_T$ region since most of the high-$p_T$ events are rejected by the jet veto.

\par
In above calculation we neglect the NLO EW correction to the $pp \rightarrow W^+ZZ \rightarrow \ell^+_1 \nu_{\ell_1} \ell^+_2 \ell^-_2 \ell^+_3 \ell^-_3 + X$ process from the real photon radiation off the final-state charged leptons. As we know, photon radiation off a final-state charged lepton reduces the lepton momentum and thus shifts many events out of the acceptance window as well as modifies the kinematic distributions of final products. This correction can be significant due to the mass-singular logarithms $\log \frac{m_{\ell}^2}{s}$ induced by the small lepton mass. We will include this contribution in leading logarithmic accuracy in future improved calculation.

\vskip 5mm
\section{Summary}
\label{sec-summary}
\par
In this paper, we investigate the NLO QCD + NLO EW corrections to the $W^{\pm}ZZ$ productions with subsequent $W^{\pm}$- and $Z$-boson leptonic decays at the $14~{\rm TeV}$ LHC, by adopting the MadSpin method which preserves both spin correlation and finite width effects to a very good accuracy. The NLO QCD+EW corrected integrated cross section and some kinematic distributions are studied. Our results demonstrate that the NLO QCD and NLO EW corrections are all significant, and modify the LO integrated cross section and some kinematic distributions obviously. In the jet-veto event selection scheme with $p_{T,jet}^{cut} = 50~ {\rm GeV}$, the NLO QCD+EW relative corrections to the integrated cross section are $20.5\%$ and $31.1\%$, while the genuine NLO EW relative corrections are $-5.42\%$ and $-4.58\%$, for the $W^+ZZ$ and $W^-ZZ$ productions, respectively. We also investigate the factorization/renormalization scale dependence of the integrated cross section, and find that the scale uncertainty is underestimated at the LO due to the absence of the strong coupling $\alpha_s$ in the LO matrix elements.

\vskip 5mm
\section{Acknowledgments}
This work was supported in part by the National Natural Science Foundation of China (No.11275190, No.11375008, No.11375171, No.11405173) and the Fundamental Research Funds for the Central Universities (No.WK2030040044).

\end{document}